\documentclass[lettersize,journal]{IEEEtran}
\usepackage{amsmath,amsfonts}
\usepackage{algorithmic}
\usepackage{algorithm}
\usepackage{array}
\usepackage[caption=false,font=normalsize,labelfont=sf,textfont=sf]{subfig}
\usepackage{textcomp}
\usepackage{stfloats}
\usepackage{url}
\usepackage{verbatim}
\usepackage{graphicx}
\usepackage{cite}
\usepackage[switch]{lineno}
\usepackage{multirow}
\usepackage{colortbl}
\usepackage{xcolor}
\usepackage{makecell}
\usepackage{hhline}
\begin{document}

\title{TEAM: Temporal Adversarial Examples Attack Model against Network Intrusion Detection System Applied to RNN}

\author{Ziyi Liu, Dengpan Ye\dag,~\IEEEmembership{Member,~IEEE,} Long Tang, Yunming Zhang and Jiacheng Deng
\thanks{This work was supported by National Natural Science Foundation of China NSFC (No. 62472325 and 62072343), the Fundamental Research Funds for the Central Universities (No. 2042023kf0228), and the National Key Research and Development Program of China (No. 2019QY(Y)0206). (\dag~Corresponding author: Dengpan Ye)}
\thanks{Ziyi Liu, Dengpan Ye, Long Tang, Yunming Zhang and Jiacheng Deng are with the Key Laboratory of Aerospace Information Security and Trusted Computing, Ministry of Education, School of Cyber Science and Engineering, Wuhan University, Wuhan, 430072, China (e-mail: ziyi\_liu@whu.edu.cn; yedp@whu.edu.cn; l\_tang@whu.edu.cn; zhangyunming@whu.deu.cn; 1462492739@qq.com).}}

\markboth{Journal of \LaTeX\ Class Files,~Vol.~14, No.~8, August~2021}%
{Shell \MakeLowercase{\textit{et al.}}: A Sample Article Using IEEEtran.cls for IEEE Journals}


\maketitle

\begin{abstract}
With the development of artificial intelligence, neural networks play a key role in network intrusion detection systems (NIDS). Despite the tremendous advantages, neural networks are susceptible to adversarial attacks. To improve the reliability of NIDS, many research has been conducted and plenty of solutions have been proposed. However, the existing solutions rarely consider the adversarial attacks against recurrent neural networks (RNN) with time steps, which would greatly affect the application of NIDS in real world. Therefore, we first propose a novel RNN adversarial attack model based on feature reconstruction called \textbf{T}emporal adversarial \textbf{E}xamples \textbf{A}ttack \textbf{M}odel \textbf{(TEAM)}, which applied to time series data and reveals the potential connection between adversarial and time steps in RNN. That is, the past adversarial examples within the same time steps can trigger further attacks on current or future original examples. Moreover, TEAM leverages Time Dilation (TD) to effectively mitigates the effect of temporal among adversarial examples within the same time steps. Experimental results show that in most attack categories, TEAM improves the misjudgment rate of NIDS on both black and white boxes, making the misjudgment rate reach more than 96.68\%. Meanwhile, the maximum increase in the misjudgment rate of the NIDS for subsequent original samples exceeds 95.57\%.
\end{abstract}


\section{Introduction}
\IEEEPARstart{W}{ith} the rapid development of modern network technology, the means of network intrusion attacks are becoming increasingly complex. The traditional static defense method~\cite{bringhenti2022automated} cannot adapt to the current complex and dynamical security requirements. To achieve active defense, Network Intrusion Detection System (NIDS)~\cite{Lei2021HNN},~\cite{Abdulganiyu2023systematic} has been widely used and achieved good results. Traditional NIDS technology achieves intrusion detection by building a knowledge base in advance and comparing the signatures in the knowledge base with the signatures extracted from the traffic~\cite{erlacher2020high}. Hence, NIDS based on traffic signatures can only detect existing attack types and is difficult to detect current complex and changeable network attacks. With the continuous development of Deep Neural Networks (DNN)~\cite{Nie2020Data}~\cite{samek2021explaining}, experts and scholars have discovered that DNN only requires to pre-train the model through a large amount of existing data to detect new intrusion attacks without the need for a priori knowledge base, which brings a new direction for the development of NIDS.

\begin{figure}
  \centering
  \includegraphics[width=0.5\textwidth]{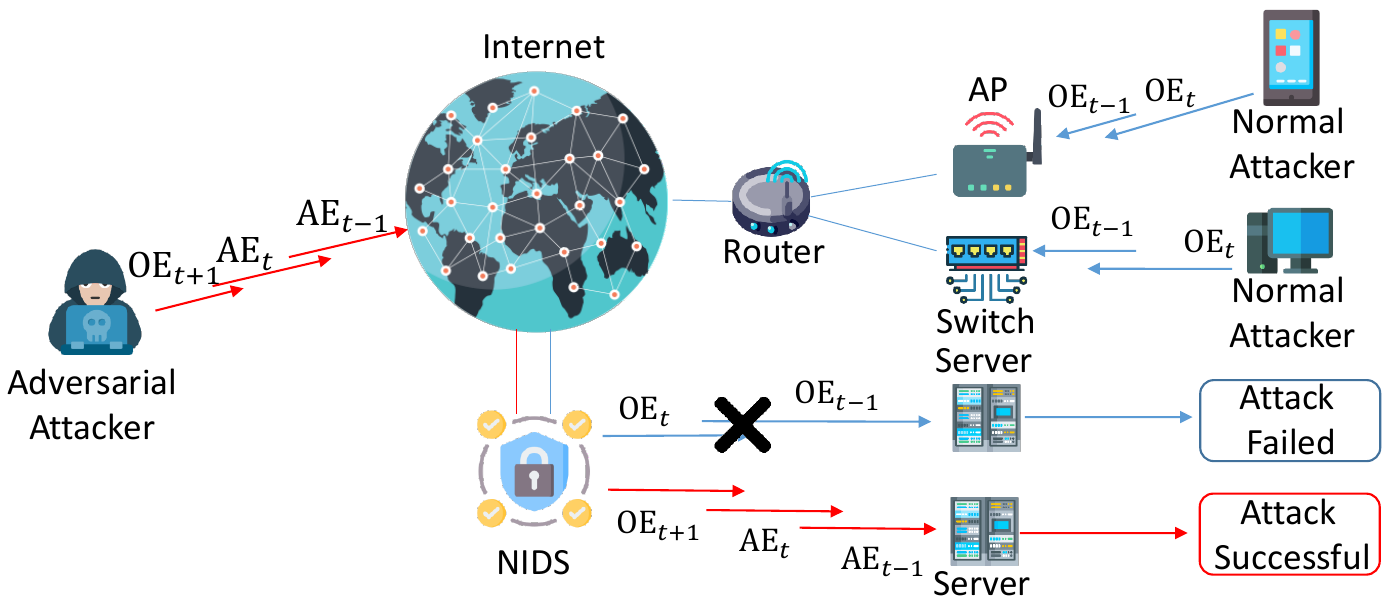}
  \caption{llustration of the TEAM attack scenario. The red part in the figure represents the use of TEAM to generate AEs to implement adversarial attacks and next moment attacks on NIDS. The blue part represents normal traffic attacks on NIDS. $AE_t$ represents the AE at time $t$ and $OE_t$ represents the OE at time $t$. It can be seen that the traditional OE attack traffic (blue) can be effectively defended by the NIDS system. However, when the attacker uses the AE traffic (red) generated by TEAM to carry out the adversarial attack, it can easily cause the NIDS to misjudge. Meanwhile, the AE traffic generated by TEAM uses the nature of the attack at the next moment to make the model misjudge the OE traffic (red) at the next moment, achieving the adversarial attack at the current moment and the adversarial attack at the next moment.}
  \label{fig:enter-label}
\end{figure}

Recurrent Neural Network (RNN)~\cite{almiani2020deep},~\cite{Kasongo2023deep}, a kind of DNN, has been widely used in NIDS because it can fully consider the temporal properties of network traffic in the characterization learning~\cite{tian2020chaotic}. Specifically, network traffic is continuously generated over time and has a temporal relationship with each other. RNN with time steps can also fully consider the impact of the previous moment data in the same time step when learning the traffic at the current moment, which greatly enhances the performance of NIDS. However, almost all DNNs have been proven to be vulnerable to Adversarial Examples (AEs)~\cite{Susilo2021Introduction},~\cite{Zhang2022Adversarial}, which brings new challenges to NIDS.

At present, a large number of researchers have conducted in-depth research on AEs attacks in NIDS, hoping to improve the defects of NIDS and enhance the robustness of NIDS~\cite{han2021evaluating}. Unfortunately, we found that there is a lack of research on adversarial attacks against RNN models with time steps in NIDS, and there are the following problems. Firstly, the RNN model is a temporal model, and traditional AEs do not consider the time characteristics of network traffic. Hence, when traditional AEs are used to attack the RNN model of NIDS, the AEs attack on the RNN model will have a large deviation due to the RNN model's learning of past moment content, resulting in a reduction of AEs attack and transferability~\cite{Alhussien2024Constraining} success rate; Secondly, in the RNN model, due to differences in network traffic structure, the time distribution of network traffic will be different. Attackers can typically only generate adversarial samples using a small portion of the network traffic dataset to simulate the data distribution. This causes the AEs have a large deviation due to the difference in time distribution when attacking the RNN model, thus reducing the attack success rate and transferability of AEs; Last but not least, because the RNN model learns to characterize part of the traffic content from past moments, the characteristics of AEs from those past moments may influence the Original Examples (OEs) in the current or future moments. This occurs due to the temporal nature of the RNN's learning process, and as a result, it could reduce the detection rate of the NIDS (OEs in this paper specifically refers to the original attack traffic in network traffic).

To solve the above problems, we first propose a model called \textbf{T}emporal adversarial \textbf{E}xamples \textbf{A}ttack \textbf{M}odel \textbf{(TEAM)} to reveal the potential connection between adversarial and time steps in RNN. In RNN models with time steps, we observe the impact of the presence of temporal between AEs within the same time step. Due to the weights~\cite{ramanujan2020s} related to past moment data in the RNN model will vary greatly depending on the data structure. Hence, targeted continuous adversarial attacks constructed by attackers using partial data have certain limitations in attacking RNN models due to the influence between AEs. In our proposed method, TEAM uses the idea of expanding the relevant weights of past moments to make the weight distribution between the attack model and the target model overlap as much as possible, effectively alleviating the above problems and realizing this type of targeted adversarial attacks. Moreover, we also observe for the first time that AEs from the past moment within the same time step have an impact on OEs at the current or future moment. Attackers can use carefully crafted AEs to cause NIDS to misjudge subsequent OEs in the same time step. This new type of attack also provides new ideas for subsequent research on the robustness of NIDS. The specific attack scenario of TEAM is shown in Figure~\ref{fig:enter-label}

Our main contributions can be summarized as follows:

$\bullet$ We systematically study adversarial attacks against RNN models with time steps in NIDS. To the best of our knowledge, we are the first to reveal potential connection between adversarial and time-stepping in such attacks. That is, past AEs in the same time step can trigger further attacks on current or future normal samples; consecutive AEs in the same time step will affect with each other, affecting the attack success rate of adversarial samples.

$\bullet$ We first propose TEAM to implement adversarial attacks against RNN models with time steps in NIDS. Meanwhile, we discovered that carefully designed AEs in an RNN model with time steps can affect the model's judgment of OEs in the same subsequent time step, and propose the concept of the next moment attack.

$\bullet$ We first propose a Time Dilation (TD) method for guided AEs generation in TEAM. The TD method adjusts the RNN model's retention of past moment content by expanding the weight of past moments, effectively alleviating the weight distribution differences in the RNN model caused by differences in data structure, thereby further improving the attack success rate of AEs.

$\bullet$ We designed a corresponding prototype system and conducted extensive experiments on the NSL-KDD dataset and the CIC-IDS2017 dataset to verify the phenomena we revealed. Experimental results show that in most attack categories, TEAM improves the misjudgment rate of NIDS on both black and white boxes, making the misjudgment rate reach more than 96.68\%. Meanwhile, in the RNN model, the next-moment attack of AE on subsequent OE generally increases the misjudgment rate of NIDS on OE, with the highest improvement reaching 95.57\%.

\section{Related Works}

\subsection{Network Intrusion Detection System}
Recently, deep learning~\cite{LeCun2015Deep} has been widely applied in NIDS. Due to the excellent representation learning capabilities, deep learning has effectively improved the detection ability of NIDS against new intrusion attacks. Mirza et al.~\cite{mirza2018computer} proposed a sequential AutoEncoder framework based on Long Short-Term Memory (LSTM) neural network to implement intrusion detection. Zhou et al.~\cite{zhou2020variational} presented an intelligent anomaly detection variational LSTM based on reconstructed feature representation for intrusion detection. Javed et al.~\cite{javed2021canintelliids} designed a new intrusion detection method called CANintelliIDS for vehicle intrusion attack detection on the CAN bus. Assis et al.~\cite{assis2021gru} proposed a SDN defense system based on the analysis of single IP traffic records, which uses the Gated Recurrent Unit (GRU) to detect DDos and intrusion attacks. Mushtaq et al.~\cite{mushtaq2022two} designed a hybrid framework including deep AutoEncoders, LSTM and Bidirectional Long Short-Term Memory (Bi-LSTM). Fu et al.\cite{fu2022Frequency} proposed a real-time malicious traffic detection system based on machine learning, which achieves high detection accuracy and high detection throughput by utilizing frequency domain features. Wang et al.~\cite{wang2023intrusion} introduced GRU into the improved AlexNet to build an intrusion detection model for urban rail transit management systems. It can be seen that the existing NIDS methods are mainly based on RNN-based models.

\subsection{Adversarial Attacks for NIDS}
Adversarial attacks have attracted widespread attention from the academic community due to their excellent attack effects on DNN-based NIDS. Yang et al.~\cite{yang2018adversarial} studied how AEs affect the performance of DNN trained to detect anomalous behavior in black-box models. Alhajjar et al.~\cite{Alhajjar2021Adversarial} explored the use of evolutionary computation (particle swarm optimization and genetic algorithms) and generative adversarial networks as tools for adversarial example generation. Clements et al.~\cite{clements2021rallying} explored the potential for adversarial entities to compromise such vulnerabilities to disrupt deep learning-based NIDS. Han et al.~\cite{han2021evaluating} used adversarial machine learning techniques to search for features located at the decision boundary of ML models, and for the first time systematically studied adversarial attacks in the gray-box/black-box traffic space. Sharon et al.~\cite{sharon2022tantra} presented a novel temporal-based end-to-end adversarial network traffic shaping attack that could bypass most NIDS. Lin et al.~\cite{lin2022idsgan} proposed a generative adversarial network framework called IDSGAN for generating adversarial malicious traffic records, aiming to attack NIDS by deceiving and evading detection. Mohammadian et al.~\cite{mohammadian2023gradient} utilized jacobian saliency maps to find the best feature groups with different features and perturbation magnitudes to generate AEs.

Although the above schemes provide a lot of research ideas for adversarial attacks against NIDS, we find that there is still a lack of systematic research on adversarial attacks in RNN models with time steps. Table~\ref{tab: relatework} summarizes the comparison of solutions related to adversarial attacks in NIDS.

\begin{table}[]

\caption{Comparison of related research on adversarial attacks in NIDS. Among them, FLA represents for Feature-Level Attacks, GFLA represents for Generative Feature-Level Attacks, and PLA represents for Packet-Level Attacks~\cite{He2023Adversarial}.}

\resizebox{\columnwidth}{!}{

\begin{tabular}{cccc}

\hline
Method     & Attack type   & \begin{tabular}[c]{@{}c@{}}RNN model \\ with temporal\end{tabular} & \begin{tabular}[c]{@{}c@{}}Next moment \\ attack\end{tabular} \\ \hline
Yang et al.~\cite{yang2018adversarial}          & FLA           & No                                                               & No                                                               \\
Alhajjar et al.~\cite{Alhajjar2021Adversarial}          & GFLA          & No                                                               & No                                                               \\
Clements et al.~\cite{clements2021rallying}          & FLA           & No                                                               & No                                                               \\
Han et al.~\cite{han2021evaluating}          & PLA           & No                                                               & No                                                               \\
Sharon et al.~\cite{sharon2022tantra}          & PLA           & No                                                               & No                                                               \\
Lin et al.~\cite{lin2022idsgan}          & GFLA          & No                                                               & No                                                               \\
Mohammadian et al.~\cite{mohammadian2023gradient}          & FLA           & No                                                               & No                                                               \\
\textbf{Ours} & \textbf{GFLA} & \textbf{Yes}                                                     & \textbf{Yes}                                                     \\ \hline
\end{tabular}}

\label{tab: relatework}
\end{table}

\section{Motivation}

\subsection{Threat Model}

We introduce a hypothetical threat scenario, \textit{i.e}., the NIDS uses an RNN to build the detection model. Note that RNN in this paper refers to the collective name of recurrent neural networks such as the Original Recurrent Neural Network (ORNN) model, the Gated Recurrent Unit (GRU) model, and the LSTM model. The attacker conducts a tentative attack on the NIDS model by accessing the network system and performing a sniffing attack~\cite{Anu2017sniffing}, accessing and monitoring the traffic, and then obtains the correct data category, traffic characteristics and approximate time step of the model from NIDS. Moreover, attackers do not know anything else about the model, such as its specific parameters, detection scheme and structure. Since the NIDS model can accurately detect anomalous traffic and process it accordingly, such as discard, purification, and other strategies, attackers want to realize targeted attacks, \textit{i.e}., to make NIDS recognize abnormal traffic as normal by adversarial means. Meanwhile, due to the data set used in our work is a network traffic feature data set (\textit{i.e}., the input data of NIDS and the data we generate AE are both "flow"). Therefore, our adversarial attacks work is also GFLA~\cite{He2023Adversarial} based on network traffic feature.

\subsection{Observations}
Network traffic is a kind of data with time continuity characteristics. Unlike traditional text data that has temporal continuity within a sentence, the temporal continuity of network traffic is reflected in the inter-traffic, such as traffic that users continue to access, continuous Dos\cite{Gao2020Detection} traffic attacks, \textit{etc}. Therefore, there exists a situation in NIDS where time steps are set in the RNN model for continuous detection of traffic data within the time steps.

The RNN model is a neural network structure that can efficiently process time series, and it can learn the data of the current moment while considering the influence of the data of the past moments on the current moment. Hence, the judgment of the current moment network traffic of NIDS based on RNN model over a period of time step not only depends on the current input traffic, but also is influenced by the traffic of the past moments.

Hence, we can easily infer that the past moment AEs within the same time step in the RNN model will affect the OEs in the current or future moments as a way to realize the next moment attack; the consecutive AEs within the same time step will affect each other, and we need to design a new mechanism to realize the normal adversarial attack in the RNN model.

\section{Methodology}
\subsection{Preliminaries}
\subsubsection{AutoEncoder}
AutoEncoder~\cite{Hinton2006Afast} is an unsupervised learning neural network model that is commonly used for data dimensionality reduction, feature learning and data reconstruction. Its basic structure includes two parts: Encoder and decoder. In this paper, we will use AutoEncoder to reconstruct data to generate AEs. The specific implementation formulas of encoder and decoder are as follows:

\begin{equation}
  h_e= A^e(W_eX+b^e),
\end{equation}

where $h^e$ represents the encoded result, $X$ represents the input of the encoder, $b^e$ represents the bias of the encoder layer, $W_e$ represents the weight of the encoder layer and $A^e$ represents the activation function of the encoder layer.

\begin{equation}
  X'= A^d(W_dh_e+b^d),
\end{equation}

where $X'$ represents the decoded result, $b^d$ represents the bias of the decoder layer, $W_d$ represents the weight of the encoder layer and $A^d$ represents the activation function of the decoder layer.

\subsubsection{Recurrent Neural Network}
Recurrent Neural Network (RNN)~\cite{Elman1990Finding} is a type of neural network with memory ability. In RNN, neurons can not only receive information at the current moment, but also from past moments. Compared with feedforward neural networks, RNNs are more consistent with the structure of biological neural networks. The existing base RNN models mainly include original RNN, LSTM~\cite{Hochreiter1997Long} and GRU~\cite{Cho2014Learning}. These base RNN models are widely used in tasks such as speech recognition, NIDS and natural language generation.

\subsubsection{Functional Features and Non-functional Features of Network Traffic}

In NIDS, network traffic has multiple attack categories and corresponding function implementations. Therefore, the selection of indicators for network traffic features is complex and diverse, which results in the features of network traffic including functional features related to current function implementation, and non-functional features that are not related to current functions. In this paper, we introduce the concepts of functional features and non-functional~\cite{lin2022idsgan} features in network traffic, whose specific meanings are as follows:

Functional features: Functional features refer to the features that affect the transmission of network traffic and the realization of specific functions (the function of a piece of network traffic is to implement Dos attacks or message transmission, \textit{etc.}) of network traffic. Modification of this type of network traffic will cause abnormalities in the protocols and functions in the network traffic, that is, the network traffic cannot be sent or received, and the original functions cannot be realized, \textit{etc.}

Non-functional features: Non-functional features refer to the features parts of network traffic that do not affect normal transmission or the realization of currently expected functions. For example, in U2R$\&$R2L\cite{Asry2023Intrusion} attack analysis of NIDS, there is almost no need to consider time-related traffic. The analysis of Dos attacks needs to consider the impact of time, but there is almost no need to consider the impact of host traffic. Therefore, for U2R$\&$R2L attacks, time-related traffic is non-functional traffic; for Dos attacks, host-related traffic is non-functional traffic.

In our approach, we only make changes to non-functional features. Therefore, even if no special adversarial constraints are imposed, the original functions of network traffic will not be affected.

\subsection{Temporal Adversarial Examples Attack Model}
\begin{figure*}[]
    \centering
        \includegraphics[width=0.8\textwidth]{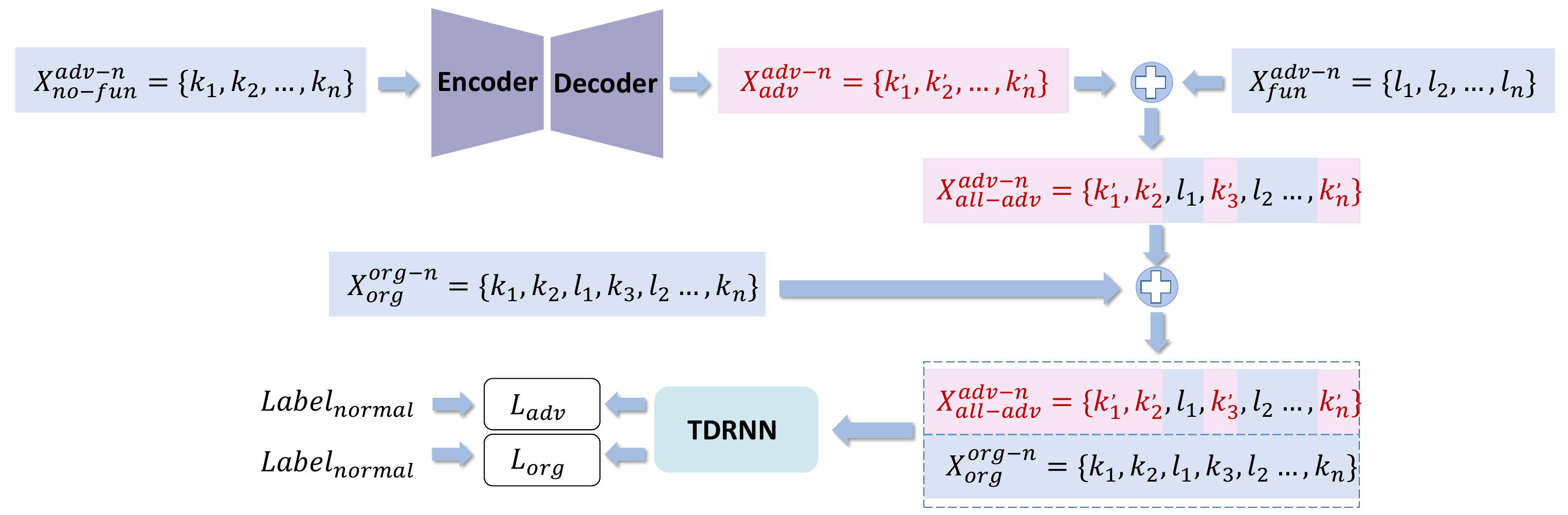}
    \caption{Illustration of the TEAM, the blue in the figure represents the unaltered OEs, and the red represents the reconstructed AEs. The process first inputs $x^{adv-n}_{no-fun}$ into AutoEncoder for data reconstruction. The reconstructed data $x^{adv-n}_{adv}$ will be spliced with the functional features $x^{adv-n}_{fun}$ to generate a new adversarial sample $x^{adv-n}_{all-adv}$. Subsequently, the $x^{adv-n}_{all-adv}$ and the normal data $x^{org-n}_{org}$ are spliced. Then, input the spliced data into the pre-trained Time Dilation RNN (TDRNN), and use the cross-entropy loss with the label to guide AutoEncoder to reconstruct the data. }
    \label{fig:pipeline}
\end{figure*}

To reveal the potential connection between confrontation and time step in RNN-based NIDS, we design a new model called TEAM, which consists of two main parts: AutoEncoder and our proposed Time Dilation RNN (TDRNN). Among them, the AutoEncoder is used to generate AEs suitable for RNN; the TDRNN is a model that has been trained to guide the generation of this AEs. Figure~\ref{fig:pipeline} illustrates the pipeline of the TEAM method, where $k_n$ represents the non-functional features of traffic, $k^{'}_n$ represents the reconstructed non-functional features, $l_n$ represents the functional features of traffic, $X^{adv-n}_{org}$ represents the original data that needs to be reconstructed into AEs at the same time step, $X^{adv-n}_{no-fun}$ represents the non-functional feature part of $X^{adv-n}_{org}$, $X^{adv-n}_{fun}$ represents the functional feature part of $X^{adv-n}_{org}$, $X^{adv-n}_{adv}$ represents AEs reconstructed by $X^{adv-n}_{no-fun}$ through AutoEncoder, $X^{adv-n}_{all-adv}$ represents network traffic with adversarial effect recomposed by $X^{adv-n}_{fun}$ and $X^{adv-n}_{adv}$, $X^{org-n}_{org}$ represents the attack traffic of original data in the same time step. This part of the symbolic content applies to Algorithm~\ref{alg:algorithm} below.

\begin{figure}[h]
    \centering
    \includegraphics[width=0.5\textwidth]{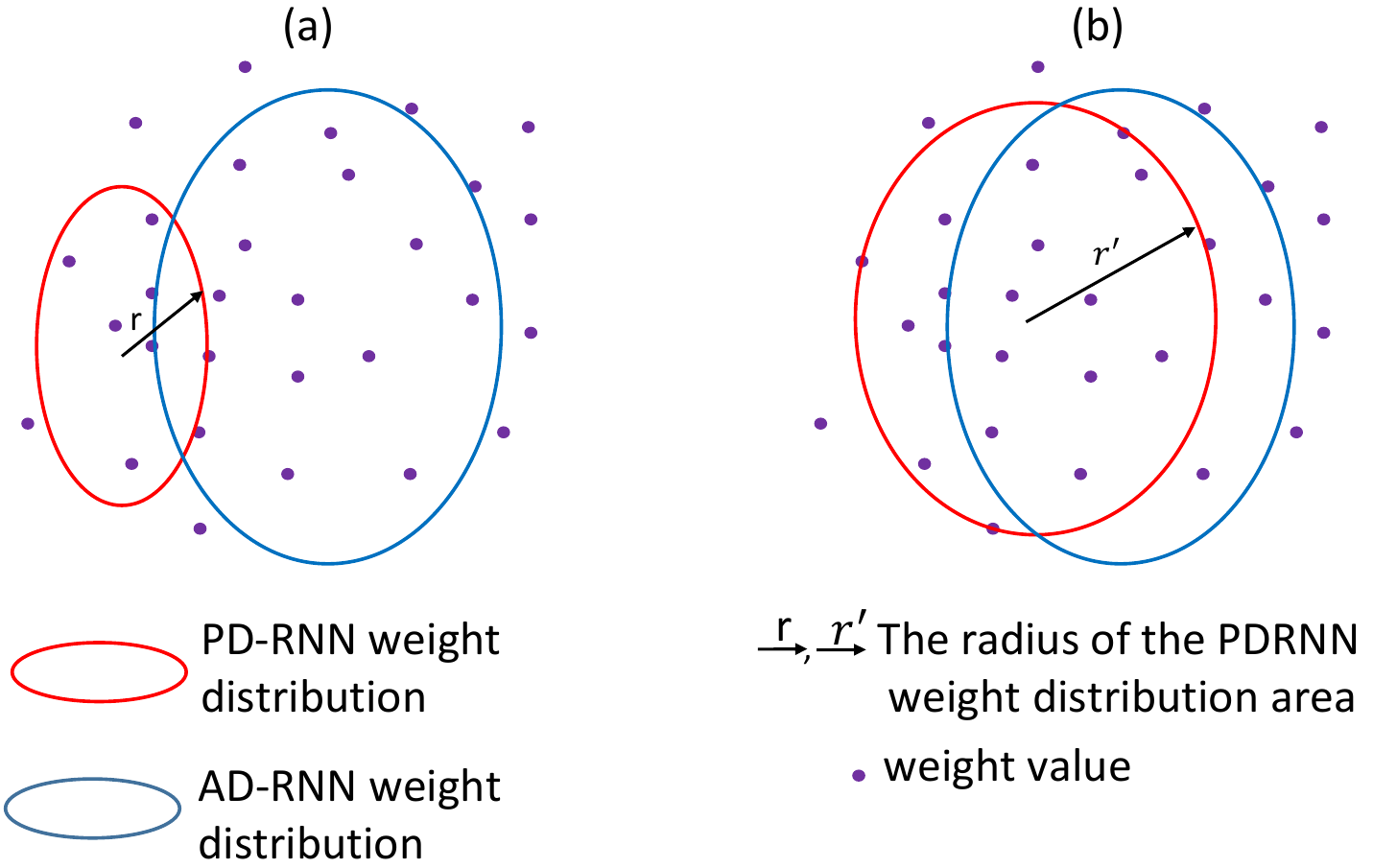}
    \caption{Illustration of Time Dilation (TD): (a) Differences in weight distribution between the PD-RNN model (without TDRNN) and the AD-RNN model. (b) Differences in weight distribution between the PD-RNN model (using TDRNN) and the AD-RNN model.}
    \label{fig:TDRNN}
\end{figure}

\subsubsection{Adversarial Attacks on RNN}
Adversarial attacks under non-RNN models can be achieved by using part of the data set to restore the feature distribution of the target model, and using similar feature distributions to guide the generation of AEs. However, the RNN model not only needs to judge the input features at the current moment in the process of realizing the detection, but also needs to consider the influence of the features in the past moment on the current moment. Hence, adversarial attacks implemented against RNN models are more complex. Particularly, we find that the AEs generated by the RNN model constructed through partial data sets (PD-RNN, the training model held by attacker) are difficult to effectively apply to the RNN model constructed through all data sets (AD-RNN, the target model that the attacker will attack), \textit{i.e}., there is an interaction between the AEs in the same time step (the current moment AEs is affected by the AEs of the past moments). This is because different data structures lead to variations in the content retained by the RNN model from past moments, and these variations cause deviations in the influence of previous data on the current moment, resulting in significant differences in the weight distributions between the PD-RNN and AD-RNN models (as shown in Figure~\ref{fig:TDRNN} (a)).

To solve the above problems, we propose an improved RNN model called Time Dilation RNN (TDRNN) to implement RNN adversarial attacks in NIDS. In TDRNN, we adjust the degree of preservation of past moment content in PD-RNN by enlarging the weights used to control past moment data, so that the weight distribution of PD-RNN and the target RNN model overlap as much as possible (as shown in Figure~\ref{fig:TDRNN} (b) shown), thereby mitigating the influence between AEs within the same time step. Our proposed TD method works on all RNN models. Below we will show the specific formulas of Time Dilation Original RNN (TDORNN), Time Dilation LSTM (TDLSTM) and Time Dilation GRU (TDGRU).

\textbf{The specific implementation formula of TDORNN is as follows:}

\begin{equation}
  h_t= tanh([x_t\cdot W_{xh}]+[h_{t-1}\cdot (W_{hh}\times hn)]),
\end{equation}

where $x_t$ represents the input at the current moment, $W_{xh}$ represents the weight input to the hidden state, $h_t$ represents the hidden state at the current moment, $h_{t-1}$ represents the hidden state at the previous moment, $W_{hh}$ represents the weight of the hidden state at the previous moment when $h_t$ is updated, and $hn$ represents the time dilation coefficient of the weight relative to the past moment

\textbf{The specific implementation formula of TDLSTM is as follows:}
\begin{equation}
  f = \sigma([x_t\cdot W_{xf}] + [h_{t-1}\cdot (W_{hf} \times hn)]),
\end{equation}

where $\sigma$ represents the Sigmoid activation function, $x_t$ represents the input at the current moment, $W_{xf}$ represents the weight input to the forget gate, $h_{t-1}$ represents the hidden state at the previous moment, $W_{hf}$ represents the weight of hidden content in the past moment in the forget gate, and $hn$ represents the time dilation coefficient of the weight relative to the past moment, and $f$ represents the forget gate, which is used to control whether to forget part of the previous cell state in LSTM.

\begin{equation}
  i = \sigma([x_t\cdot W_{xi}] + [h_{t-1}\cdot (W_{hi} \times hn)]),
\end{equation}

\begin{equation}
  c = tanh([x_t\cdot (W_{xc}\times hn)] + [h_{t-1}\cdot (W_{hc} \times hn)]),
\end{equation}

\begin{equation}
  cc = f*cc+i*c,
\end{equation}

where $W_{xf}$ represents the weight input to the input gate, $W_{hf}$ represents the weight of hidden content in the past moment in the input gate, $W_{xc}$ represents the weight input to the cell candidate status, $W_{hc}$ represents the weight of hidden content in the past moment in the cell candidate status, $i$ represents the input gate, which is used to decide which new information in LSTM should be added to the cell state, $c$ represents the candidate value for updating cell state in LSTM, and $cc$ represents the cell state of the current time step.

\begin{equation}
  o = \sigma([x_t\cdot (W_{xo} \times hn)] + [h_{t-1}\cdot (W_{ho} \times hn)]),
\end{equation}

\begin{equation}
  h_t = o * tanh(cc),
\end{equation}

where $W_{xf}$ represents the weight input to the output gate, $W_{ho}$ represents the weight of hidden content in the past moment in the output gate, $o$ is used to control the hidden state to be output in LSTM, and $h_t$ represents the hidden state at the current moment.

\textbf{The specific implementation formula of TDGRU is as follows:}

\begin{equation}
  R_t=\sigma([x_t\cdot W_{xr}]+[h_{t-1}\cdot (W_{hr}\times hn)]),
\end{equation}

\begin{equation}
  Z_t=\sigma([x_t\cdot W_{xz}]+[h_{t-1}\cdot (W_{hz}\times hn)]),
\end{equation}
where $\sigma$ represents the Sigmoid activation function, $R_t$ and $Z_t$ represent reset gate and update gate, $x_t$ represents the input at the current moment, $h_t$ represents the hidden state at the current moment, $h_{t-1}$ represents the hidden state at the past moment, $W_{xr}$ and $W_{xz}$ represent the weight assigned by $x_t$ in the reset gate and update gate respectively, $W_{hr}$ and $W_{hz}$ represent the weight assigned by $h_t$ in the reset gate and update gate respectively, and $hn$ represents the time dilation coefficient of the weight relative to the past moment.

\begin{equation}
  \tilde{h_t} = tanh([x_t\cdot (W_{xh}\times hn)] + [(r\times h_{t-1})\cdot (W_{hh} \times hn)]),
\end{equation}
where $\tilde{h_t}$ represents the candidate state, $W_{xh}$ represents the weight assigned to $x_t$ in the candidate state, and $W_{hh}$ represents the weight assigned to $h_{t-1}$ in the candidate state. Note that we also add a TD coefficient to the weight of $x_t$ in $\tilde{h_t}$, which is to balance the ratio of current moment data and past moment data in the candidate set.

\begin{equation}
  h_t = Z_t\times \tilde{h_t}+(1-Z_t)\times h_{t-1},
\end{equation}
where $h_t$ represents the currently hidden state.

It is worth noting that in our proposed method, the above $hn$ is used as the weight time dilation coefficient of the past time in the RNN to participate in the training phase of the model, and is still retained in the attack phase after the training.

Because TDRNN is used to guide the generation of AutoEncoders during training, we use the cross-entropy loss function as a condition for generating AEs during AutoEncoder training. It is worth noting that the functional characteristics in network traffic will affect the specific function implementation of network traffic. In our method, we feed non-functional features of network traffic into an AutoEncoder for adversarial example generation. Non-functional changes will not have a major impact on the functionality of network traffic. Therefore, we believe that there is no need to add too many constraints to the changes in this part. The specific loss function of this part is as follows:

\begin{equation}
  L_{adv} = CrossEntropy(x^{n}_{adv}, Label_{normal}).
\end{equation}

\subsubsection{Next Moment Attack in RNN Model}
Since RNN can fully consider the time impact between network traffic, we find that AEs can be used in RNN to change the model's judgment of OEs at subsequent moments in the same time step. Specifically, attackers only need to send a small number of consecutive AEs to influence the model's judgment on most traffic in the same time step, which greatly affects the reliability and robustness of the NIDS model. Meanwhile, it also gives attackers greater operating space and lower costs. We use TEAM to implement this type of attack. Specifically, we only need to splice the OEs behind the AEs generated by the AutoEncoder when training the RNN AEs in the same time step. Then, Time Dilation RNN (TDRNN) is used to conduct guided training on the spliced data, where the loss function given for the OE part is as follows.
\begin{equation}
  L_{org} = CrossEntropy(x^{n}_{org}, Label_{normal}).
\end{equation}

We use this loss function to make the generated past moment AEs influence the current or future OEs as much as possible to achieve the next moment attack. Therefore, by combining with the loss function when generating adversarial samples in the previous section, we can know that the total loss function of TEAM is as follows.
\begin{equation}
  L_{CrossEntropy} = L_{adv} + L_{org}.
\end{equation}

The specific training and implementation algorithms for adversarial attacks and the next attack on the RNN model are shown in Algorithm~\ref{alg:algorithm}. In Algorithm~\ref{alg:algorithm}, we use $x$ and $\hat{x}$ to represent the training set and test set in the overall data $X$, respectively.

\begin{algorithm}[tb]
    \caption{Adversarial attack and next moment attack in RNN model}
    \label{alg:algorithm}
    \textbf{Input}: $X^{adv-n}_{org}$, $X^{adv-n}_{no-fun}$, $X^{adv-n}_{fun}$, $X^{adv-n}_{adv}$, $X^{adv-n}_{all-adv}$, $X^{org-n}_{org}$ \\
    \textbf{Parameter}: $time-n$ (length of time step), $adv-n$ (the number of AEs used to generate in a time step), $org-n$ (the number of original samples in the time step), $epoch-n$ (number of iteration cycles)\\
    \textbf{Output}: $\hat{x}^{adv-n}_{adv}$
    \begin{algorithmic}[1] 
        \STATE Extract non-functional features: $x^{adv-n}_{org}\rightarrow x^{adv-n}_{no-fun}$.

        \FOR{$epoch-i$ in $epoch-n$}
        \STATE  $x^{adv-n}_{adv}\leftarrow AutoEncoder(x^{adv-n}_{no-fun})$.
        \STATE  $x^{adv-n}_{all-adv}\leftarrow cat(x^{adv-n}_{fun}, x^{adv-n}_{adv})$.
        \STATE  $x_{time-n} \leftarrow cat(x^{adv-n}_{adv}, x^{org-n}_{org}) $.
        \STATE  $label\leftarrow TDRNN(x_time-n)$.
        \STATE  $min(Loss_{CrossEntropy}(label, label_{Normal}))$.
        \ENDFOR
        \STATE Extract non-functional features: $\hat{x}^{adv-n}_{org}\rightarrow \hat{x}^{adv-n}_{no-fun}$.
        \STATE $\hat{x}^{adv-n}_{adv} \leftarrow AutoEncoder(\hat{x}^{adv-n}_{no-fun})$.
        \STATE $\hat{x}^{adv-n}_{all-adv}\leftarrow cat(\hat{x}^{adv-n}_{fun}, \hat{x}^{adv-n}_{adv})$.
        \RETURN $\hat{x}^{adv-n}_{adv}$.

    \end{algorithmic}
\end{algorithm}

\section{Experiments}
\subsection{Experimental Settings}
\subsubsection{Datasets}
In this paper, we use NSL-KDD~\cite{tavallaee2009detailed} and CIC-IDS2017~\cite{Sharafaldin2017Toward} datasets as experimental datasets.

\textbf{NSL-KDD:} As one of the most widely used benchmark datasets in NIDS, the NSL-KDD dataset contains 125,973 training samples and 22,544 test samples, with four different attack types. We refer to the existing NIDS adversarial attack literature, make a detailed division of the functional and non-functional features of the NSL-KDD dataset~\cite{lin2022idsgan}. In network attacks, attackers can use sniffing attacks to obtain part of the network access traffic. Therefore, we use half of the training set for model training (\textit{i.e.} PD-RNN) to generate AE, and use the full training set to train the target NIDS (\textit{i.e.} AD-RNN) to test AE and next-moment attacks.

\textbf{CIC-IDS2017:} The CIC-IDS2017 dataset is often used in NIDS detection experiments because it reflects the characteristics of modern network traffic. It contains 12 traffics of different attack types. Each traffic in the official CSV file for machine learning consists of 78 features. The entire dataset is collected from Monday to Friday. In this experiment, we conducted a targeted adversarial attack to make NIDS misjudge the attack traffic as normal. Due to all the data collected by the official on Monday is normal data, this experiment excludes the data of that day from the dataset. Therefore, in this experiment, the training set is divided into 1,802,513 and the test set is 339,599. Similar to NSL-KDD, we use half of the training set for model training (\textit{i.e.} PD-RNN) to generate AE, and use the full training set to train the target NIDS (\textit{i.e.} AD-RNN) to test AE and the next moment attack. In order to more effectively verify the effectiveness of our proposed method, this half of the CIC-IDS2017 dataset is shuffled, which has a higher randomness of temporal sequence. Meanwhile, to ensure the data balance of the experimental data as much as possible, we merge all Dos attacks in CIC-IDS2017 into one Dos attack. Infiltration and Botnet both involve infiltrating the network, hiding activities, maintaining infection for a long time and performing malicious operations. Therefore, these two are merged into one Infiltration$\&$Botnet attack, and SSH-Patator and FTP-Patator types are merged into Patator attacks. In addition, since there are only 11 traffic in Heartbleed attack types and it is difficult to merge them with other types, we do not consider this attack type in this experiment.

We refer to the official literature of CIC-IDS2017~\cite{Sharafaldin2017Toward} and divide the functional and non-functional features of traffic in combination with the properties of each attack type. For example, in Dos traffic, we need to exclude those features that directly involve core traffic patterns such as traffic magnitude, packet size, and time interval. It is worth noting that we directly classify the features related to the flag bit into the functional features, because such features are usually only discrete specific values. Imposing disturbances on such features will cause excessive damage or reduce the concealment of the attack. For example, mistakenly changing the FIN or RST flag bit may cause the connection to be closed or reset in advance. In addition, as far as we know, there is no clear division between the non-functional and functional features of the attack types of CIC-IDS2017 in the existing literature. Most of them only divide the weight of the features in different attacks of CIC-IDS2017 (such as literature~\cite{Msika2019Sigma}~\cite{Hoang2022DA}). Therefore, in the division process, we only take the feature types that are irrelevant to the functional implementation of this type of attack. We acknowledge that this may cause some non-functional features to be missed, but adding perturbations to fewer non-functional features to achieve adversarial attacks can better illustrate the effectiveness of our proposed method. The attack types and corresponding non-functional feature selections in this experiment are shown in Table~\ref{tab:nonfeature}.

\begin{table}[]
\setlength{\arrayrulewidth}{0.3mm}
\caption{Attack types and non-functional features of the CIC-IDS2017 dataset in this experiment.}
\renewcommand
\arraystretch{1.2}  
\resizebox{\linewidth}{!}{
\begin{tabular}{c|l|c|l}
\hline
\multirow{2}{*}{\begin{tabular}[c]{@{}c@{}}Network\\ traffic type\end{tabular}} & \multicolumn{1}{c|}{\multirow{2}{*}{\begin{tabular}[c]{@{}c@{}}Non-functional \\ features\end{tabular}}}                                                                                                                                                                                                                                                                         & \multirow{2}{*}{\begin{tabular}[c]{@{}c@{}}Network\\ traffic type\end{tabular}}    & \multicolumn{1}{c}{\multirow{2}{*}{\begin{tabular}[c]{@{}c@{}}Non-functional \\ features\end{tabular}}}                                                                                                                                                                                                                                                                                               \\
                                                                                & \multicolumn{1}{c|}{}                                                                                                                                                                                                                                                                                                                                                            &                                                                                    & \multicolumn{1}{c}{}                                                                                                                                                                                                                                                                                                                                                                                  \\ \hline
\multirow{7}{*}{Dos}                                                            & \multirow{7}{*}{\begin{tabular}[c]{@{}l@{}}Subflow Fwd Packets, \\ Subflow Fwd Bytes, \\ Subflow Bwd Packets, \\ Subflow Bwd Bytes, \\ Active Std, Active Max, \\ Idle Mean, Idle Std, Idle Max, Idle Min\end{tabular}}                                                                                                                                                          & \multirow{7}{*}{\begin{tabular}[c]{@{}c@{}}Infiltration\\ \\ $\&$Botnet\end{tabular}} & \multirow{7}{*}{\begin{tabular}[c]{@{}l@{}}Flow Bytes/s, Flow Packets/s, \\ Flow IAT Std, Fwd IAT Std, Bwd IAT Std, \\ Fwd Header Length, Bwd Header Length, \\ Packet Length Std, Packet Length Variance, \\ Fwd Header Length.1, Subflow Fwd Packets, \\ Subflow Fwd Bytes, Subflow Bwd Packets, \\ Subflow Bwd Bytes, Active Std, Idle Std\end{tabular}}                                           \\
                                                                                &                                                                                                                                                                                                                                                                                                                                                                                  &                                                                                    &                                                                                                                                                                                                                                                                                                                                                                                                       \\
                                                                                &                                                                                                                                                                                                                                                                                                                                                                                  &                                                                                    &                                                                                                                                                                                                                                                                                                                                                                                                       \\
                                                                                &                                                                                                                                                                                                                                                                                                                                                                                  &                                                                                    &                                                                                                                                                                                                                                                                                                                                                                                                       \\
                                                                                &                                                                                                                                                                                                                                                                                                                                                                                  &                                                                                    &                                                                                                                                                                                                                                                                                                                                                                                                       \\
                                                                                &                                                                                                                                                                                                                                                                                                                                                                                  &                                                                                    &                                                                                                                                                                                                                                                                                                                                                                                                       \\
                                                                                &                                                                                                                                                                                                                                                                                                                                                                                  &                                                                                    &                                                                                                                                                                                                                                                                                                                                                                                                       \\ \hline
\multirow{8}{*}{Patator}                                                        & \multirow{8}{*}{\begin{tabular}[c]{@{}l@{}}Fwd Avg Bytes/Bulk, Fwd Avg Packets/Bulk, \\ Fwd Avg Bulk Rate, Bwd Avg Bytes/Bulk, \\ Bwd Avg Packets/Bulk, Bwd Avg Bulk Rate, \\ Subflow Fwd Packets, Subflow Fwd Bytes, \\ Subflow Bwd Packets, Subflow Bwd Bytes, \\ Active Mean, Active Std, Active Max, \\ Active Min, Idle Mean, Idle Std, \\ Idle Max, Idle Min\end{tabular}} & \multirow{8}{*}{Portscan}                                                          & \multirow{8}{*}{\begin{tabular}[c]{@{}l@{}}Packet Length Mean, Packet Length Std, \\ Packet Length Variance, Down/Up Ratio, \\ Average Packet Size, Avg Fwd Segment Size, \\ Avg Bwd Segment Size, Subflow Fwd Packets, \\ Subflow Fwd Bytes, Subflow Bwd Packets, \\ Subflow Bwd Bytes, Active Mean, Active Std, \\ Active Max, Active Min, Idle Mean, Idle Std, \\ Idle Max, Idle Min\end{tabular}} \\
                                                                                &                                                                                                                                                                                                                                                                                                                                                                                  &                                                                                    &                                                                                                                                                                                                                                                                                                                                                                                                       \\
                                                                                &                                                                                                                                                                                                                                                                                                                                                                                  &                                                                                    &                                                                                                                                                                                                                                                                                                                                                                                                       \\
                                                                                &                                                                                                                                                                                                                                                                                                                                                                                  &                                                                                    &                                                                                                                                                                                                                                                                                                                                                                                                       \\
                                                                                &                                                                                                                                                                                                                                                                                                                                                                                  &                                                                                    &                                                                                                                                                                                                                                                                                                                                                                                                       \\
                                                                                &                                                                                                                                                                                                                                                                                                                                                                                  &                                                                                    &                                                                                                                                                                                                                                                                                                                                                                                                       \\
                                                                                &                                                                                                                                                                                                                                                                                                                                                                                  &                                                                                    &                                                                                                                                                                                                                                                                                                                                                                                                       \\
                                                                                &                                                                                                                                                                                                                                                                                                                                                                                  &                                                                                    &                                                                                                                                                                                                                                                                                                                                                                                                       \\ \hline
\multirow{7}{*}{Web Attack}                                                     & \multirow{7}{*}{\begin{tabular}[c]{@{}l@{}}Fwd Packet Length Mean, \\ Bwd Packet Length Mean, \\ Min Packet Length, Max Packet Length, \\ Packet Length Mean, Packet Length Std, \\ Packet Length Variance, Down/Up Ratio, \\ Average Packet Size\end{tabular}}                                                                                                                  & \multirow{7}{*}{DDos}                                                              & \multirow{7}{*}{\begin{tabular}[c]{@{}l@{}}Down/Up Ratio, Subflow Fwd Packets, \\ Subflow Fwd Bytes, Subflow Bwd Packets, \\ Subflow Bwd Bytes, Active Mean, Active Std, \\ Active Max, Active Min, Idle Mean, Idle Std, \\ Idle Max, Idle Min\end{tabular}}                                                                                                                                          \\
                                                                                &                                                                                                                                                                                                                                                                                                                                                                                  &                                                                                    &                                                                                                                                                                                                                                                                                                                                                                                                       \\
                                                                                &                                                                                                                                                                                                                                                                                                                                                                                  &                                                                                    &                                                                                                                                                                                                                                                                                                                                                                                                       \\
                                                                                &                                                                                                                                                                                                                                                                                                                                                                                  &                                                                                    &                                                                                                                                                                                                                                                                                                                                                                                                       \\
                                                                                &                                                                                                                                                                                                                                                                                                                                                                                  &                                                                                    &                                                                                                                                                                                                                                                                                                                                                                                                       \\
                                                                                &                                                                                                                                                                                                                                                                                                                                                                                  &                                                                                    &                                                                                                                                                                                                                                                                                                                                                                                                       \\
                                                                                &                                                                                                                                                                                                                                                                                                                                                                                  &                                                                                    &                                                                                                                                                                                                                                                                                                                                                                                                       \\ \hline
\end{tabular}}
\label{tab:nonfeature}
\end{table}

\subsubsection{Compare Models and Evaluation Metrics}
To the best of our knowledge, ours is the first work to investigate the potential connection between adversarial and time stepping of RNN models in NIDS. In addition, because the RNN model with time steps will consider the impact of past moment traffic on current moment traffic when characterization learning, other existing work on adversarial attacks against NIDS difficult to be directly adapted to RNN models with time steps. Based on the above reasons, in this paper, we will compare the existing typical adversarial attack methods PGD~\cite{Madrye2017Towards}, C$\&$W~\cite{Carlini2017Towards}, IDSGAN~\cite{lin2022idsgan} and J-Attack~\cite{mohammadian2023gradient} to verify the adversarial and transferability of our proposed method.

In our experiments, we implemented a targeted attack, use Attack Success Rate (ASR) as an evaluation metric, use Misjudgment Accuracy Rate (MAR) to represent the probability that normal attack traffic in NIDS is misjudged as normal traffic, use MAR1 to represent the probability that NIDS misjudged the OE attack traffic at the first moment after AEs as normal traffic in the original model, use MAR2 to represent the probability that NIDS misjudged the OE attack traffic at the second moment after AEs as normal traffic in the original model. Note that our adversarial attack is a targeted adversarial attack. When the AE of a certain attack category in the test set is identified as normal traffic, we can consider that the ASR of the adversarial attack is the same as the MSR of this attack category by NIDS. Moreover, in the table of experimental results, we bold the optimal data of each item. If the data in a category are all optimal values, we bold the data of our proposed method.

\subsubsection{Experimental Model and Details}
We conduct adversarial attacks and next time step attacks on three existing baseline RNN models (ORNN, LSTM and GRU) respectively. Meanwhile, we also perform black-box transferability between models. Furthermore, all our attacks are targeted attacks (making NIDS identify abnormal traffic as normal traffic). In the NSL-KDD data set, exceptions are the U2R$\&$R2L attack categories, the functional features and non-functional feature areas of the other attack categories are different. Therefore, different AEs will be generated according to categories for experiments.

For the NSL-KDD dataset, the number of network iteration epochs used by TEAM to generate AE is 100, the gradient optimization method is Adam, and the learning rate is 0.001. For the CIC-IDS2017 dataset, the maximum number of network iteration epochs used by TEAM to generate AE is 2000, and we will take the better AE generated in the epoch cycle, the gradient optimization method is Adam, and the learning rate is 0.001. Meanwhile, the intermediate layer of AutoEncoder is 4, and the time step of the RNN model is set to 8. In the same time step, the first 6 samples are AEs and the remaining 2 are OEs.

\begin{table}[]
\setlength{\arrayrulewidth}{0.3mm}
\caption{Adversarial attacks on RNN using the NSL-KDD dataset (gray background represents white-box attacks, white background represents black-box transferability attacks). }
\renewcommand
\arraystretch{1.2}
\resizebox{\columnwidth}{!}{
\begin{tabular}{c|c|c|cc|cc|cc}
\hline
&                                                                             &                          & \multicolumn{2}{c|}{Original RNN}                                                      & \multicolumn{2}{c|}{LSTM}                                                              & \multicolumn{2}{c}{GRU}                                                                \\ \cline{4-9}
\multirow{-2}{*}{\begin{tabular}[c]{@{}c@{}}Network\\ traffic type\end{tabular}} & \multirow{-2}{*}{\begin{tabular}[c]{@{}c@{}}IDS\\ Model\end{tabular}}       & \multirow{-2}{*}{Method} & \multicolumn{1}{c|}{MAR(\%)}                  & ASR(\%)                                & \multicolumn{1}{c|}{MAR(\%)}                  & ASR(\%)                                & \multicolumn{1}{c|}{MAR(\%)}                  & ASR(\%)                                \\ \hline
&                                                                             & PGD                      & \multicolumn{1}{c|}{}                         & \cellcolor[HTML]{C0C0C0}26.2           & \multicolumn{1}{c|}{}                         & 52.77                                  & \multicolumn{1}{c|}{}                         & 22.63                                  \\
&                                                                             & C$\&$W                   & \multicolumn{1}{c|}{}                         & \cellcolor[HTML]{C0C0C0}97.31          & \multicolumn{1}{c|}{}                         & 34.51                                  & \multicolumn{1}{c|}{}                         & 89.53                                  \\
&                                                                             & IDSGAN                   & \multicolumn{1}{c|}{}                         & \cellcolor[HTML]{C0C0C0}76.82          & \multicolumn{1}{c|}{}                         & 73.51                                  & \multicolumn{1}{c|}{}                         & 60.15                                  \\
&                                                                             & J-Attack                 & \multicolumn{1}{c|}{}                         & \cellcolor[HTML]{C0C0C0}0              & \multicolumn{1}{c|}{}                         & 18.2                                   & \multicolumn{1}{c|}{}                         & 0                                      \\
& \multirow{-5}{*}{\begin{tabular}[c]{@{}c@{}}Original\\ \\ RNN\end{tabular}} & TEAM(ours)               & \multicolumn{1}{c|}{}                         & \cellcolor[HTML]{C0C0C0}\textbf{100}   & \multicolumn{1}{c|}{}                         & \textbf{100}                           & \multicolumn{1}{c|}{}                         & \textbf{100}                           \\ \hhline{|~|-|-|~|-|~|-|~|-|}
&                                                                             & PGD                      & \multicolumn{1}{c|}{}                         & 92.34                                  & \multicolumn{1}{c|}{}                         & \cellcolor[HTML]{C0C0C0}72.51          & \multicolumn{1}{c|}{}                         & 87.01                                  \\
&                                                                             & C$\&$W                   & \multicolumn{1}{c|}{}                         & 91.23                                  & \multicolumn{1}{c|}{}                         & \cellcolor[HTML]{C0C0C0}37.33          & \multicolumn{1}{c|}{}                         & 87.44                                  \\
&                                                                             & IDSGAN                   & \multicolumn{1}{c|}{}                         & 29.45                                  & \multicolumn{1}{c|}{}                         & \cellcolor[HTML]{C0C0C0}17.2           & \multicolumn{1}{c|}{}                         & 18.41                                  \\
&                                                                             & J-Attack                 & \multicolumn{1}{c|}{}                         & 0                                      & \multicolumn{1}{c|}{}                         & \cellcolor[HTML]{C0C0C0}18.2           & \multicolumn{1}{c|}{}                         & 0                                      \\
& \multirow{-5}{*}{LSTM}                                                      & TEAM(ours)               & \multicolumn{1}{c|}{}                         & \textbf{100}                           & \multicolumn{1}{c|}{}                         & \cellcolor[HTML]{C0C0C0}\textbf{100}   & \multicolumn{1}{c|}{}                         & \textbf{100}                           \\ \hhline{|~|-|-|~|-|~|-|~|-|}
&                                                                             & PGD                      & \multicolumn{1}{c|}{}                         & 87.01                                  & \multicolumn{1}{c|}{}                         & 77.59                                  & \multicolumn{1}{c|}{}                         & \cellcolor[HTML]{C0C0C0}81.33          \\
&                                                                             & C$\&$W                   & \multicolumn{1}{c|}{}                         & 53                                     & \multicolumn{1}{c|}{}                         & 25.03                                  & \multicolumn{1}{c|}{}                         & \cellcolor[HTML]{C0C0C0}50.94          \\
&                                                                             & IDSGAN                   & \multicolumn{1}{c|}{}                         & 59.99                                  & \multicolumn{1}{c|}{}                         & 30.84                                  & \multicolumn{1}{c|}{}                         & \cellcolor[HTML]{C0C0C0}23.17          \\
&                                                                             & J-Attack                 & \multicolumn{1}{c|}{}                         & 0                                      & \multicolumn{1}{c|}{}                         & 18.2                                   & \multicolumn{1}{c|}{}                         & \cellcolor[HTML]{C0C0C0}0              \\
\multirow{-15}{*}{Dos}                                                           & \multirow{-5}{*}{GRU}                                                       & TEAM(ours)               & \multicolumn{1}{c|}{\multirow{-15}{*}{22.86}} & \textbf{100}                           & \multicolumn{1}{c|}{\multirow{-15}{*}{23.74}} & \textbf{100}                           & \multicolumn{1}{c|}{\multirow{-15}{*}{21.7}}  & \cellcolor[HTML]{C0C0C0}\textbf{100}   \\ \hline
&                                                                             & PGD                      & \multicolumn{1}{c|}{}                         & \cellcolor[HTML]{C0C0C0}99.18          & \multicolumn{1}{c|}{}                         & 97.28                                  & \multicolumn{1}{c|}{}                         & 86.4                                   \\
&                                                                             & C$\&$W                   & \multicolumn{1}{c|}{}                         & \cellcolor[HTML]{C0C0C0}96.16          & \multicolumn{1}{c|}{}                         & 96.2                                   & \multicolumn{1}{c|}{}                         & 69.19                                  \\
&                                                                             & IDSGAN                   & \multicolumn{1}{c|}{}                         & \cellcolor[HTML]{C0C0C0}78.63          & \multicolumn{1}{c|}{}                         & 98.37                                  & \multicolumn{1}{c|}{}                         & 86.63                                  \\
&                                                                             & J-Attack                 & \multicolumn{1}{c|}{}                         & \cellcolor[HTML]{C0C0C0}65.76          & \multicolumn{1}{c|}{}                         & 97.1                                   & \multicolumn{1}{c|}{}                         & 32.7                                   \\
& \multirow{-5}{*}{\begin{tabular}[c]{@{}c@{}}Original\\ \\ RNN\end{tabular}} & TEAM(ours)               & \multicolumn{1}{c|}{}                         & \cellcolor[HTML]{C0C0C0}\textbf{99.77} & \multicolumn{1}{c|}{}                         & \textbf{99.81}                         & \multicolumn{1}{c|}{}                         & \textbf{97.65}                         \\ \hhline{|~|-|-|~|-|~|-|~|-|}
&                                                                             & PGD                      & \multicolumn{1}{c|}{}                         & 92.34                                  & \multicolumn{1}{c|}{}                         & \cellcolor[HTML]{C0C0C0}72.51          & \multicolumn{1}{c|}{}                         & 87.64                                  \\
&                                                                             & C$\&$W                   & \multicolumn{1}{c|}{}                         & 96.56                                  & \multicolumn{1}{c|}{}                         & \cellcolor[HTML]{C0C0C0}96.79          & \multicolumn{1}{c|}{}                         & 70.23                                  \\
&                                                                             & IDSGAN                   & \multicolumn{1}{c|}{}                         & 89.65                                  & \multicolumn{1}{c|}{}                         & \cellcolor[HTML]{C0C0C0}97.24          & \multicolumn{1}{c|}{}                         & 82.88                                  \\
&                                                                             & J-Attack                 & \multicolumn{1}{c|}{}                         & 65.76                                  & \multicolumn{1}{c|}{}                         & \cellcolor[HTML]{C0C0C0}97.1           & \multicolumn{1}{c|}{}                         & 32.7                                   \\
& \multirow{-5}{*}{LSTM}                                                      & TEAM(ours)               & \multicolumn{1}{c|}{}                         & \textbf{99.81}                         & \multicolumn{1}{c|}{}                         & \cellcolor[HTML]{C0C0C0}\textbf{99.86} & \multicolumn{1}{c|}{}                         & \textbf{99.81}                         \\ \hhline{|~|-|-|~|-|~|-|~|-|}
&                                                                             & PGD                      & \multicolumn{1}{c|}{}                         & 99.72                                  & \multicolumn{1}{c|}{}                         & 97.77                                  & \multicolumn{1}{c|}{}                         & \cellcolor[HTML]{C0C0C0}99.41          \\
&                                                                             & C$\&$W                   & \multicolumn{1}{c|}{}                         & 98.32                                  & \multicolumn{1}{c|}{}                         & 99.09                                  & \multicolumn{1}{c|}{}                         & \cellcolor[HTML]{C0C0C0}70.05          \\
&                                                                             & IDSGAN                   & \multicolumn{1}{c|}{}                         & 97.92                                  & \multicolumn{1}{c|}{}                         & 99.5                                   & \multicolumn{1}{c|}{}                         & \cellcolor[HTML]{C0C0C0}85.45          \\
&                                                                             & J-Attack                 & \multicolumn{1}{c|}{}                         & 65.76                                  & \multicolumn{1}{c|}{}                         & 97.1                                   & \multicolumn{1}{c|}{}                         & \cellcolor[HTML]{C0C0C0}32.7           \\
\multirow{-15}{*}{U2R$\&$R2L}                                                    & \multirow{-5}{*}{GRU}                                                       & TEAM(ours)               & \multicolumn{1}{c|}{\multirow{-15}{*}{91.29}} & \textbf{99.77}                         & \multicolumn{1}{c|}{\multirow{-15}{*}{95.76}} & \textbf{99.9}                          & \multicolumn{1}{c|}{\multirow{-15}{*}{91.09}} & \cellcolor[HTML]{C0C0C0}\textbf{99.86} \\ \hline
&                                                                             & PGD                      & \multicolumn{1}{c|}{}                         & \cellcolor[HTML]{C0C0C0}62.52          & \multicolumn{1}{c|}{}                         & 66.72                                  & \multicolumn{1}{c|}{}                         & 66.55                                  \\
&                                                                             & C$\&$W                   & \multicolumn{1}{c|}{}                         & \cellcolor[HTML]{C0C0C0}58.71          & \multicolumn{1}{c|}{}                         & 50.66                                  & \multicolumn{1}{c|}{}                         & 58.33                                  \\
&                                                                             & IDSGAN                   & \multicolumn{1}{c|}{}                         & \cellcolor[HTML]{C0C0C0}23.73          & \multicolumn{1}{c|}{}                         & 13.85                                  & \multicolumn{1}{c|}{}                         & 27.42                                  \\
&                                                                             & J-Attack                 & \multicolumn{1}{c|}{}                         & \cellcolor[HTML]{C0C0C0}33.44          & \multicolumn{1}{c|}{}                         & 32.06                                  & \multicolumn{1}{c|}{}                         & 33.49                                  \\
& \multirow{-5}{*}{\begin{tabular}[c]{@{}c@{}}Original\\ \\ RNN\end{tabular}} & TEAM(ours)               & \multicolumn{1}{c|}{}                         & \cellcolor[HTML]{C0C0C0}\textbf{100}   & \multicolumn{1}{c|}{}                         & \textbf{100}                           & \multicolumn{1}{c|}{}                         & \textbf{100}                           \\ \hhline{|~|-|-|~|-|~|-|~|-|}
&                                                                             & PGD                      & \multicolumn{1}{c|}{}                         & 84.16                                  & \multicolumn{1}{c|}{}                         & \cellcolor[HTML]{C0C0C0}69.97          & \multicolumn{1}{c|}{}                         & 80.73                                  \\
&                                                                             & C$\&$W                   & \multicolumn{1}{c|}{}                         & 67.1                                   & \multicolumn{1}{c|}{}                         & \cellcolor[HTML]{C0C0C0}64.4           & \multicolumn{1}{c|}{}                         & 65.61                                  \\
&                                                                             & IDSGAN                   & \multicolumn{1}{c|}{}                         & 49.66                                  & \multicolumn{1}{c|}{}                         & \cellcolor[HTML]{C0C0C0}46.19          & \multicolumn{1}{c|}{}                         & 47.07                                  \\
&                                                                             & J-Attack                 & \multicolumn{1}{c|}{}                         & 33.44                                  & \multicolumn{1}{c|}{}                         & \cellcolor[HTML]{C0C0C0}32.06          & \multicolumn{1}{c|}{}                         & 33.49                                  \\
& \multirow{-5}{*}{LSTM}                                                      & TEAM(ours)               & \multicolumn{1}{c|}{}                         & \textbf{99.94}                         & \multicolumn{1}{c|}{}                         & \cellcolor[HTML]{C0C0C0}\textbf{100}   & \multicolumn{1}{c|}{}                         & \textbf{100}                           \\ \hhline{|~|-|-|~|-|~|-|~|-|}
&                                                                             & PGD                      & \multicolumn{1}{c|}{}                         & 91.55                                  & \multicolumn{1}{c|}{}                         & 95.19                                  & \multicolumn{1}{c|}{}                         & \cellcolor[HTML]{C0C0C0}98.01          \\
&                                                                             & C$\&$W                   & \multicolumn{1}{c|}{}                         & 56.34                                  & \multicolumn{1}{c|}{}                         & 56.18                                  & \multicolumn{1}{c|}{}                         & \cellcolor[HTML]{C0C0C0}58.11          \\
&                                                                             & IDSGAN                   & \multicolumn{1}{c|}{}                         & 1.6                                    & \multicolumn{1}{c|}{}                         & 3.36                                   & \multicolumn{1}{c|}{}                         & \cellcolor[HTML]{C0C0C0}0.71           \\
&                                                                             & J-Attack                 & \multicolumn{1}{c|}{}                         & 33.44                                  & \multicolumn{1}{c|}{}                         & 32.06                                  & \multicolumn{1}{c|}{}                         & \cellcolor[HTML]{C0C0C0}33.49          \\
\multirow{-15}{*}{Probe}                                                         & \multirow{-5}{*}{GRU}                                                       & TEAM(ours)               & \multicolumn{1}{c|}{\multirow{-15}{*}{39.87}} & \textbf{100}                           & \multicolumn{1}{c|}{\multirow{-15}{*}{18.38}} & \textbf{100}                           & \multicolumn{1}{c|}{\multirow{-15}{*}{36.65}} & \cellcolor[HTML]{C0C0C0}\textbf{100}   \\ \hline
\end{tabular}}
\label{tab:AEattack}
\end{table}

\subsection{Results and Evaluations}
In this section, we compare the differences between our proposed method and other four typical RNN model methods in terms of AE white-box attack success rate and black-box transferability. Ablation experiments were conducted to verify the effectiveness of our proposed TDRNN method. At the same time, we experimentally verified the concept of next-moment attack. Since our proposed method uses the first 6 OEs for AE in each time step, and the last two OEs are used to verify the concept of attack at the next moment. Therefore, in this experiment, we will take the same number of AEs for comparison and verification.

\subsubsection{Adversarial Attacks on RNNs Using the NSL-KDD Dataset}

In this section, we first conduct adversarial attacks on RNN-based NIDS on the NSL-KDD dataset to verify the effectiveness of the proposed method. The experimental results are shown in Table~\ref{tab:AEattack}.

From Table~\ref{tab:AEattack}, we can see that in the NSL-KDD dataset, the TEAM method we proposed better guides the generation of AEs, and the attack success rate and transferability of the generated AEs on RNN, LSTM and GRU are significantly better than those of other comparison methods. The reason is the AE generated by the PGD and C$\&$W methods is affected by the difference in the weight distribution due to data differences between the PD-RNN model and the target RNN model, cannot accurately implement adversarial attacks against the target RNN. In the IDSGAN method, due to the generator and discriminator use traditional fully connected neural networks, the PD-RNN model only serves as a query for generating results. Therefore, the AE generated by IDSGAN is difficult to adapt to the temporal of RNN-type models. J-Attack is limited by the scope of feature selection (that is, it can only select perturbation to add targets in non-functional features), and its method of directly using masks to add perturbations is also difficult to effectively deal with the timing differences between RNNs. The TEAM method we proposed not only guides AE generation through the RNN model, but also uses the time dilation method we proposed to minimize the difference in weight distribution between the PD-RNN model and the target RNN model, and then the generated AE can utilize the temporal of the model itself in adversarial attacks as much as possible to achieve effective AE white-box attacks and black-box transferability attacks.

We can also see that IDSGAN has extremely low ASR in Dos and Probe\cite{Tan2021Invisible}. We observed the experimental results and found that the RNN classifier misclassified AE as other attack types. For example, the vast majority of Probe attack AEs are predicted to be Dos attacks by the RNN classifier. This is because in attack types with slightly higher temporal such as Probe, the AE generated by IDSGAN will cross the boundary excessively because the RNN model retains past content, affecting the normal judgment of the RNN model and causing the attack to fail.

\begin{table*}[]
\setlength{\arrayrulewidth}{0.3mm}
\caption{Adversarial attacks on RNN using the CIC-IDS2017 dataset (gray background represents white-box attacks, white background represents black-box transferability attacks). }
\renewcommand
\arraystretch{1.2}  
\resizebox{\linewidth}{!}{
\begin{tabular}{c|c|c|cc|cc|cc|c|c|c|cc|cc|cc}
\hline
&                                                                             &                          & \multicolumn{2}{c|}{Original RNN}                                                      & \multicolumn{2}{c|}{LSTM}                                                              & \multicolumn{2}{c|}{GRU}                                                             &                                                                                   &                                                                             &                          & \multicolumn{2}{c|}{Original RNN}                                                      & \multicolumn{2}{c|}{LSTM}                                                              & \multicolumn{2}{c}{GRU}                                                              \\ 
\hhline{~|~|~|-|-|-|-|-|-|~|~|~|-|-|-|-|-|-}
\multirow{-2}{*}{\begin{tabular}[c]{@{}c@{}}Network\\ traffic type\end{tabular}} & \multirow{-2}{*}{\begin{tabular}[c]{@{}c@{}}IDS\\ Model\end{tabular}}       & \multirow{-2}{*}{Method} & \multicolumn{1}{c|}{MAR(\%)}                  & ASR(\%)                                & \multicolumn{1}{c|}{MAR(\%)}                  & ASR(\%)                                & \multicolumn{1}{c|}{MAR(\%)}                  & ASR(\%)                              & \multirow{-2}{*}{\begin{tabular}[c]{@{}c@{}}Network\\ traffic type\end{tabular}}  & \multirow{-2}{*}{\begin{tabular}[c]{@{}c@{}}IDS\\ Model\end{tabular}}       & \multirow{-2}{*}{Method} & \multicolumn{1}{c|}{MAR(\%)}                  & ASR(\%)                                & \multicolumn{1}{c|}{MAR(\%)}                  & ASR(\%)                                & \multicolumn{1}{c|}{ASR(\%)}                  & MAR(\%)                              \\ \hline
&                                                                             & PGD                      & \multicolumn{1}{c|}{}                         & \cellcolor[HTML]{C0C0C0}85.55          & \multicolumn{1}{c|}{}                         & 99.25                                  & \multicolumn{1}{c|}{}                         & 98.31                                &                                                                                   &                                                                             & PGD                      & \multicolumn{1}{c|}{}                         & \cellcolor[HTML]{C0C0C0}55.81          & \multicolumn{1}{c|}{}                         & 98.84                                  & \multicolumn{1}{c|}{}                         & 93.02                                \\
&                                                                             & C$\&$W                   & \multicolumn{1}{c|}{}                         & \cellcolor[HTML]{C0C0C0}99.87          & \multicolumn{1}{c|}{}                         & 97.21                                  & \multicolumn{1}{c|}{}                         & 100                                  &                                                                                   &                                                                             & C$\&$W                   & \multicolumn{1}{c|}{}                         & \cellcolor[HTML]{C0C0C0}66.57          & \multicolumn{1}{c|}{}                         & 94.77                                  & \multicolumn{1}{c|}{}                         & 100                                  \\
&                                                                             & IDSGAN                   & \multicolumn{1}{c|}{}                         & \cellcolor[HTML]{C0C0C0}98.42          & \multicolumn{1}{c|}{}                         & 99.56                                  & \multicolumn{1}{c|}{}                         & 99.13                                &                                                                                   &                                                                             & IDSGAN                   & \multicolumn{1}{c|}{}                         & \cellcolor[HTML]{C0C0C0}83.33          & \multicolumn{1}{c|}{}                         & 97.28                                  & \multicolumn{1}{c|}{}                         & 90.69                                \\
&                                                                             & J-Attack                 & \multicolumn{1}{c|}{}                         & \cellcolor[HTML]{C0C0C0}95.65          & \multicolumn{1}{c|}{}                         & 99.13                                  & \multicolumn{1}{c|}{}                         & 99.21                                &                                                                                   &                                                                             & J-Attack                 & \multicolumn{1}{c|}{}                         & \cellcolor[HTML]{C0C0C0}84.49          & \multicolumn{1}{c|}{}                         & 96.89                                  & \multicolumn{1}{c|}{}                         & 87.98                                \\
& \multirow{-5}{*}{\begin{tabular}[c]{@{}c@{}}Original\\ \\ RNN\end{tabular}} & TEAM(ours)               & \multicolumn{1}{c|}{}                         & \cellcolor[HTML]{C0C0C0}\textbf{99.97} & \multicolumn{1}{c|}{}                         & \textbf{99.8}                          & \multicolumn{1}{c|}{}                         & \textbf{100}                         &                                                                                   & \multirow{-5}{*}{\begin{tabular}[c]{@{}c@{}}Original\\ \\ RNN\end{tabular}} & TEAM(ours)               & \multicolumn{1}{c|}{}                         & \cellcolor[HTML]{C0C0C0}\textbf{97.67} & \multicolumn{1}{c|}{}                         & \textbf{100}                           & \multicolumn{1}{c|}{}                         & \textbf{100}                         \\
\hhline{~|-|-|~|-|~|-|~|-|~|-|-|~|-|~|-|~|-}
&                                                                             & PGD                      & \multicolumn{1}{c|}{}                         & 89.47                                  & \multicolumn{1}{c|}{}                         & \cellcolor[HTML]{C0C0C0}99.25          & \multicolumn{1}{c|}{}                         & 98.31                                &                                                                                   &                                                                             & PGD                      & \multicolumn{1}{c|}{}                         & 82.85                                  & \multicolumn{1}{c|}{}                         & \cellcolor[HTML]{C0C0C0}98.84          & \multicolumn{1}{c|}{}                         & 97.38                                \\
&                                                                             & C$\&$W                   & \multicolumn{1}{c|}{}                         & 99.78                                  & \multicolumn{1}{c|}{}                         & \cellcolor[HTML]{C0C0C0}97.21          & \multicolumn{1}{c|}{}                         & 100                                  &                                                                                   &                                                                             & C$\&$W                   & \multicolumn{1}{c|}{}                         & 78.2                                   & \multicolumn{1}{c|}{}                         & \cellcolor[HTML]{C0C0C0}94.19          & \multicolumn{1}{c|}{}                         & 99.13                                \\
&                                                                             & IDSGAN                   & \multicolumn{1}{c|}{}                         & 95.3                                   & \multicolumn{1}{c|}{}                         & \cellcolor[HTML]{C0C0C0}96.82          & \multicolumn{1}{c|}{}                         & 94.76                                &                                                                                   &                                                                             & IDSGAN                   & \multicolumn{1}{c|}{}                         & 85.27                                  & \multicolumn{1}{c|}{}                         & \cellcolor[HTML]{C0C0C0}97.28          & \multicolumn{1}{c|}{}                         & 94.18                                \\
&                                                                             & J-Attack                 & \multicolumn{1}{c|}{}                         & 95.65                                  & \multicolumn{1}{c|}{}                         & \cellcolor[HTML]{C0C0C0}99.13          & \multicolumn{1}{c|}{}                         & 99.21                                &                                                                                   &                                                                             & J-Attack                 & \multicolumn{1}{c|}{}                         & 84.49                                  & \multicolumn{1}{c|}{}                         & \cellcolor[HTML]{C0C0C0}96.89          & \multicolumn{1}{c|}{}                         & 87.98                                \\
& \multirow{-5}{*}{LSTM}                                                      & TEAM(ours)               & \multicolumn{1}{c|}{}                         & \textbf{100}                           & \multicolumn{1}{c|}{}                         & \cellcolor[HTML]{C0C0C0}\textbf{99.94} & \multicolumn{1}{c|}{}                         & \textbf{100}                         &                                                                                   & \multirow{-5}{*}{LSTM}                                                      & TEAM(ours)               & \multicolumn{1}{c|}{}                         & \textbf{98.44}                         & \multicolumn{1}{c|}{}                         & \cellcolor[HTML]{C0C0C0}\textbf{100}   & \multicolumn{1}{c|}{}                         & \textbf{100}                         \\ 
\hhline{~|-|-|~|-|~|-|~|-|~|-|-|~|-|~|-|~|-}
&                                                                             & PGD                      & \multicolumn{1}{c|}{}                         & 80.09                                  & \multicolumn{1}{c|}{}                         & 98.56                                  & \multicolumn{1}{c|}{}                         & \cellcolor[HTML]{C0C0C0}97.58        &                                                                                   &                                                                             & PGD                      & \multicolumn{1}{c|}{}                         & 65.99                                  & \multicolumn{1}{c|}{}                         & 98.55                                  & \multicolumn{1}{c|}{}                         & \cellcolor[HTML]{C0C0C0}97.38        \\
&                                                                             & C$\&$W                   & \multicolumn{1}{c|}{}                         & 98                                     & \multicolumn{1}{c|}{}                         & 97.33                                  & \multicolumn{1}{c|}{}                         & \cellcolor[HTML]{C0C0C0}99.02        &                                                                                   &                                                                             & C$\&$W                   & \multicolumn{1}{c|}{}                         & 77.03                                  & \multicolumn{1}{c|}{}                         & 95.06                                  & \multicolumn{1}{c|}{}                         & \cellcolor[HTML]{C0C0C0}93.31        \\
&                                                                             & IDSGAN                   & \multicolumn{1}{c|}{}                         & 98.91                                  & \multicolumn{1}{c|}{}                         & 99.7                                   & \multicolumn{1}{c|}{}                         & \cellcolor[HTML]{C0C0C0}99.18        &                                                                                   &                                                                             & IDSGAN                   & \multicolumn{1}{c|}{}                         & 91.08                                  & \multicolumn{1}{c|}{}                         & 98.83                                  & \multicolumn{1}{c|}{}                         & \cellcolor[HTML]{C0C0C0}93.41        \\
&                                                                             & J-Attack                 & \multicolumn{1}{c|}{}                         & 95.65                                  & \multicolumn{1}{c|}{}                         & 99.13                                  & \multicolumn{1}{c|}{}                         & \cellcolor[HTML]{C0C0C0}99.21        &                                                                                   &                                                                             & J-Attack                 & \multicolumn{1}{c|}{}                         & 84.49                                  & \multicolumn{1}{c|}{}                         & 96.89                                  & \multicolumn{1}{c|}{}                         & \cellcolor[HTML]{C0C0C0}87.98        \\
\multirow{-15}{*}{Dos}                                                           & \multirow{-5}{*}{GRU}                                                       & TEAM(ours)               & \multicolumn{1}{c|}{\multirow{-15}{*}{24.09}} & \textbf{100}                           & \multicolumn{1}{c|}{\multirow{-15}{*}{9.41}}  & \textbf{99.97}                         & \multicolumn{1}{c|}{\multirow{-15}{*}{2.72}}  & \cellcolor[HTML]{C0C0C0}\textbf{100} & \multirow{-15}{*}{\begin{tabular}[c]{@{}c@{}}Infiltration\\ $\&$Botnet\end{tabular}} & \multirow{-5}{*}{GRU}                                                       & TEAM(ours)               & \multicolumn{1}{c|}{\multirow{-15}{*}{30.52}} & \textbf{100}                           & \multicolumn{1}{c|}{\multirow{-15}{*}{63.08}} & \textbf{100}                           & \multicolumn{1}{c|}{\multirow{-15}{*}{55.52}} & \cellcolor[HTML]{C0C0C0}\textbf{100} \\ \hline
&                                                                             & PGD                      & \multicolumn{1}{c|}{}                         & \cellcolor[HTML]{C0C0C0}77.47          & \multicolumn{1}{c|}{}                         & 98.31                                  & \multicolumn{1}{c|}{}                         & 97.79                                &                                                                                   &                                                                             & PGD                      & \multicolumn{1}{c|}{}                         & \cellcolor[HTML]{C0C0C0}52.93          & \multicolumn{1}{c|}{}                         & 88.56                                  & \multicolumn{1}{c|}{}                         & 84.48                                \\
&                                                                             & C$\&$W                   & \multicolumn{1}{c|}{}                         & \cellcolor[HTML]{C0C0C0}66.8           & \multicolumn{1}{c|}{}                         & 97.01                                  & \multicolumn{1}{c|}{}                         & 99.87                                &                                                                                   &                                                                             & C$\&$W                   & \multicolumn{1}{c|}{}                         & \cellcolor[HTML]{C0C0C0}33.92          & \multicolumn{1}{c|}{}                         & 99.9                                   & \multicolumn{1}{c|}{}                         & 100                                  \\
&                                                                             & IDSGAN                   & \multicolumn{1}{c|}{}                         & \cellcolor[HTML]{C0C0C0}99.82          & \multicolumn{1}{c|}{}                         & 100                                    & \multicolumn{1}{c|}{}                         & 100                                  &                                                                                   &                                                                             & IDSGAN                   & \multicolumn{1}{c|}{}                         & \cellcolor[HTML]{C0C0C0}42.03          & \multicolumn{1}{c|}{}                         & 97.81                                  & \multicolumn{1}{c|}{}                         & 89.96                                \\
&                                                                             & J-Attack                 & \multicolumn{1}{c|}{}                         & \cellcolor[HTML]{C0C0C0}97.22          & \multicolumn{1}{c|}{}                         & 99.13                                  & \multicolumn{1}{c|}{}                         & 96.7                                 &                                                                                   &                                                                             & J-Attack                 & \multicolumn{1}{c|}{}                         & \cellcolor[HTML]{C0C0C0}80.77          & \multicolumn{1}{c|}{}                         & 78.4                                   & \multicolumn{1}{c|}{}                         & 67.98                                \\
 & \multirow{-5}{*}{\begin{tabular}[c]{@{}c@{}}Original\\ \\ RNN\end{tabular}} & TEAM(ours)               & \multicolumn{1}{c|}{}                         & \cellcolor[HTML]{C0C0C0}\textbf{100}   & \multicolumn{1}{c|}{}                         & \textbf{100}                           & \multicolumn{1}{c|}{}                         & \textbf{100}                         &                                                                                   & \multirow{-5}{*}{\begin{tabular}[c]{@{}c@{}}Original\\ \\ RNN\end{tabular}} & TEAM(ours)               & \multicolumn{1}{c|}{}                         & \cellcolor[HTML]{C0C0C0}\textbf{100}   & \multicolumn{1}{c|}{}                         & \textbf{99.9}                          & \multicolumn{1}{c|}{}                         & \textbf{100}                         \\
 \hhline{~|-|-|~|-|~|-|~|-|~|-|-|~|-|~|-|~|-}
 &                                                                             & PGD                      & \multicolumn{1}{c|}{}                         & 69.66                                  & \multicolumn{1}{c|}{}                         & \cellcolor[HTML]{C0C0C0}96.74          & \multicolumn{1}{c|}{}                         & 98.7                                 &                                                                                   &                                                                             & PGD                      & \multicolumn{1}{c|}{}                         & 26.83                                  & \multicolumn{1}{c|}{}                         & \cellcolor[HTML]{C0C0C0}94.58          & \multicolumn{1}{c|}{}                         & 84.95                                \\
 &                                                                             & C$\&$W                   & \multicolumn{1}{c|}{}                         & 66.41                                  & \multicolumn{1}{c|}{}                         & \cellcolor[HTML]{C0C0C0}97.01          & \multicolumn{1}{c|}{}                         & 100                                  &                                                                                   &                                                                             & C$\&$W                   & \multicolumn{1}{c|}{}                         & 37.83                                  & \multicolumn{1}{c|}{}                         & \cellcolor[HTML]{C0C0C0}99.9           & \multicolumn{1}{c|}{}                         & 100                                  \\
 &                                                                             & IDSGAN                   & \multicolumn{1}{c|}{}                         & 97.91                                  & \multicolumn{1}{c|}{}                         & \cellcolor[HTML]{C0C0C0}\textbf{100}   & \multicolumn{1}{c|}{}                         & 99.65                                &                                                                                   &                                                                             & IDSGAN                   & \multicolumn{1}{c|}{}                         & 32.06                                  & \multicolumn{1}{c|}{}                         & \cellcolor[HTML]{C0C0C0}98.23          & \multicolumn{1}{c|}{}                         & 99.91                                \\
 &                                                                             & J-Attack                 & \multicolumn{1}{c|}{}                         & 97.22                                  & \multicolumn{1}{c|}{}                         & \cellcolor[HTML]{C0C0C0}99.47          & \multicolumn{1}{c|}{}                         & 96.7                                 &                                                                                   &                                                                             & J-Attack                 & \multicolumn{1}{c|}{}                         & 80.77                                  & \multicolumn{1}{c|}{}                         & \cellcolor[HTML]{C0C0C0}78.4           & \multicolumn{1}{c|}{}                         & 67.98                                \\
 & \multirow{-5}{*}{LSTM}                                                      & TEAM(ours)               & \multicolumn{1}{c|}{}                         & \textbf{100}                           & \multicolumn{1}{c|}{}                         & \cellcolor[HTML]{C0C0C0}99.82          & \multicolumn{1}{c|}{}                         & \textbf{100}                         &                                                                                   & \multirow{-5}{*}{LSTM}                                                      & TEAM(ours)               & \multicolumn{1}{c|}{}                         & \textbf{96.68}                         & \multicolumn{1}{c|}{}                         & \cellcolor[HTML]{C0C0C0}\textbf{99.93} & \multicolumn{1}{c|}{}                         & \textbf{100}                         \\ 
  \hhline{~|-|-|~|-|~|-|~|-|~|-|-|~|-|~|-|~|-}
 &                                                                             & PGD                      & \multicolumn{1}{c|}{}                         & 61.07                                  & \multicolumn{1}{c|}{}                         & 97.79                                  & \multicolumn{1}{c|}{}                         & \cellcolor[HTML]{C0C0C0}98.18        &                                                                                   &                                                                             & PGD                      & \multicolumn{1}{c|}{}                         & 26.83                                  & \multicolumn{1}{c|}{}                         & 94.58                                  & \multicolumn{1}{c|}{}                         & \cellcolor[HTML]{C0C0C0}84.95        \\
 &                                                                             & C$\&$W                   & \multicolumn{1}{c|}{}                         & 96.88                                  & \multicolumn{1}{c|}{}                         & 96.48                                  & \multicolumn{1}{c|}{}                         & \cellcolor[HTML]{C0C0C0}98.05        &                                                                                   &                                                                             & C$\&$W                   & \multicolumn{1}{c|}{}                         & 37.83                                  & \multicolumn{1}{c|}{}                         & 99.9                                   & \multicolumn{1}{c|}{}                         & \cellcolor[HTML]{C0C0C0}100          \\
 &                                                                             & IDSGAN                   & \multicolumn{1}{c|}{}                         & 99.65                                  & \multicolumn{1}{c|}{}                         & 100                                    & \multicolumn{1}{c|}{}                         & \cellcolor[HTML]{C0C0C0}99.82        &                                                                                   &                                                                             & IDSGAN                   & \multicolumn{1}{c|}{}                         & 98.88                                  & \multicolumn{1}{c|}{}                         & 99.97                                  & \multicolumn{1}{c|}{}                         & \cellcolor[HTML]{C0C0C0}100          \\
 &                                                                             & J-Attack                 & \multicolumn{1}{c|}{}                         & 97.22                                  & \multicolumn{1}{c|}{}                         & 99.13                                  & \multicolumn{1}{c|}{}                         & \cellcolor[HTML]{C0C0C0}96.7         &                                                                                   &                                                                             & J-Attack                 & \multicolumn{1}{c|}{}                         & 80.77                                  & \multicolumn{1}{c|}{}                         & 78.4                                   & \multicolumn{1}{c|}{}                         & \cellcolor[HTML]{C0C0C0}67.98        \\
\multirow{-15}{*}{Patator}                                                       & \multirow{-5}{*}{GRU}                                                       & TEAM(ours)               & \multicolumn{1}{c|}{\multirow{-15}{*}{77.46}} & \textbf{100}                           & \multicolumn{1}{c|}{\multirow{-15}{*}{57.38}} & \textbf{100}                           & \multicolumn{1}{c|}{\multirow{-15}{*}{24.35}} & \cellcolor[HTML]{C0C0C0}\textbf{100} & \multirow{-15}{*}{Portscan}                                                       & \multirow{-5}{*}{GRU}                                                       & TEAM(ours)               & \multicolumn{1}{c|}{\multirow{-15}{*}{0.5}}   & \textbf{100}                           & \multicolumn{1}{c|}{\multirow{-15}{*}{2.06}}  & \textbf{99.98}                         & \multicolumn{1}{c|}{\multirow{-15}{*}{9.4}}   & \cellcolor[HTML]{C0C0C0}\textbf{100} \\ \hline
&                                                                             & PGD                      & \multicolumn{1}{c|}{}                         & \cellcolor[HTML]{C0C0C0}88.54          & \multicolumn{1}{c|}{}                         & 81.25                                  & \multicolumn{1}{c|}{}                         & 57.64                                &                                                                                   &                                                                             & PGD                      & \multicolumn{1}{c|}{}                         & \cellcolor[HTML]{C0C0C0}53.61          & \multicolumn{1}{c|}{}                         & 94.69                                  & \multicolumn{1}{c|}{}                         & 93.1                                 \\
&                                                                             & C$\&$W                   & \multicolumn{1}{c|}{}                         & \cellcolor[HTML]{C0C0C0}97.22          & \multicolumn{1}{c|}{}                         & 97.92                                  & \multicolumn{1}{c|}{}                         & 82.99                                &                                                                                   &                                                                             & C$\&$W                   & \multicolumn{1}{c|}{}                         & \cellcolor[HTML]{C0C0C0}40.08          & \multicolumn{1}{c|}{}                         & 99.68                                  & \multicolumn{1}{c|}{}                         & 100                                  \\
&                                                                             & IDSGAN                   & \multicolumn{1}{c|}{}                         & \cellcolor[HTML]{C0C0C0}83.79          & \multicolumn{1}{c|}{}                         & 97.68                                  & \multicolumn{1}{c|}{}                         & 85.18                                &                                                                                   &                                                                             & IDSGAN                   & \multicolumn{1}{c|}{}                         & \cellcolor[HTML]{C0C0C0}49.7           & \multicolumn{1}{c|}{}                         & 62.28                                  & \multicolumn{1}{c|}{}                         & 61.61                                \\
&                                                                             & J-Attack                 & \multicolumn{1}{c|}{}                         & \cellcolor[HTML]{C0C0C0}100            & \multicolumn{1}{c|}{}                         & 98.61                                  & \multicolumn{1}{c|}{}                         & 96.29                                &                                                                                   &                                                                             & J-Attack                 & \multicolumn{1}{c|}{}                         & \cellcolor[HTML]{C0C0C0}43.1           & \multicolumn{1}{c|}{}                         & 32.91                                  & \multicolumn{1}{c|}{}                         & 35.62                                \\
& \multirow{-5}{*}{\begin{tabular}[c]{@{}c@{}}Original\\ \\ RNN\end{tabular}} & TEAM(ours)               & \multicolumn{1}{c|}{}                         & \cellcolor[HTML]{C0C0C0}\textbf{100}   & \multicolumn{1}{c|}{}                         & \textbf{100}                           & \multicolumn{1}{c|}{}                         & \textbf{100}                         &                                                                                   & \multirow{-5}{*}{\begin{tabular}[c]{@{}c@{}}Original\\ \\ RNN\end{tabular}} & TEAM(ours)               & \multicolumn{1}{c|}{}                         & \cellcolor[HTML]{C0C0C0}\textbf{100}   & \multicolumn{1}{c|}{}                         & \textbf{100}                           & \multicolumn{1}{c|}{}                         & \textbf{100}                         \\
 \hhline{~|-|-|~|-|~|-|~|-|~|-|-|~|-|~|-|~|-}
&                                                                             & PGD                      & \multicolumn{1}{c|}{}                         & 90.28                                  & \multicolumn{1}{c|}{}                         & \cellcolor[HTML]{C0C0C0}94.79          & \multicolumn{1}{c|}{}                         & 75                                   &                                                                                   &                                                                             & PGD                      & \multicolumn{1}{c|}{}                         & 58.07                                  & \multicolumn{1}{c|}{}                         & \cellcolor[HTML]{C0C0C0}96.11          & \multicolumn{1}{c|}{}                         & 96.68                                \\
&                                                                             & C$\&$W                   & \multicolumn{1}{c|}{}                         & 94.1                                   & \multicolumn{1}{c|}{}                         & \cellcolor[HTML]{C0C0C0}97.22          & \multicolumn{1}{c|}{}                         & 84.38                                &                                                                                   &                                                                             & C$\&$W                   & \multicolumn{1}{c|}{}                         & 40.82                                  & \multicolumn{1}{c|}{}                         & \cellcolor[HTML]{C0C0C0}99.68          & \multicolumn{1}{c|}{}                         & 99.85                                \\
&                                                                             & IDSGAN                   & \multicolumn{1}{c|}{}                         & 80.55                                  & \multicolumn{1}{c|}{}                         & \cellcolor[HTML]{C0C0C0}97.68          & \multicolumn{1}{c|}{}                         & 85.18                                &                                                                                   &                                                                             & IDSGAN                   & \multicolumn{1}{c|}{}                         & 0.08                                   & \multicolumn{1}{c|}{}                         & \cellcolor[HTML]{C0C0C0}1.41           & \multicolumn{1}{c|}{}                         & 1.11                                 \\
&                                                                             & J-Attack                 & \multicolumn{1}{c|}{}                         & 100                                    & \multicolumn{1}{c|}{}                         & \cellcolor[HTML]{C0C0C0}98.61          & \multicolumn{1}{c|}{}                         & 33.49                                &                                                                                   &                                                                             & J-Attack                 & \multicolumn{1}{c|}{}                         & 43.1                                   & \multicolumn{1}{c|}{}                         & \cellcolor[HTML]{C0C0C0}32.91          & \multicolumn{1}{c|}{}                         & 35.62                                \\
& \multirow{-5}{*}{LSTM}                                                      & TEAM(ours)               & \multicolumn{1}{c|}{}                         & \textbf{100}                           & \multicolumn{1}{c|}{}                         & \cellcolor[HTML]{C0C0C0}\textbf{100}   & \multicolumn{1}{c|}{}                         & \textbf{100}                         &                                                                                   & \multirow{-5}{*}{LSTM}                                                      & TEAM(ours)               & \multicolumn{1}{c|}{}                         & \textbf{100}                           & \multicolumn{1}{c|}{}                         & \cellcolor[HTML]{C0C0C0}\textbf{99.84} & \multicolumn{1}{c|}{}                         & \textbf{100}                         \\ 
 \hhline{~|-|-|~|-|~|-|~|-|~|-|-|~|-|~|-|~|-}
&                                                                             & PGD                      & \multicolumn{1}{c|}{}                         & 86.46                                  & \multicolumn{1}{c|}{}                         & 86.46                                  & \multicolumn{1}{c|}{}                         & \cellcolor[HTML]{C0C0C0}57.99        &                                                                                   &                                                                             & PGD                      & \multicolumn{1}{c|}{}                         & 36.12                                  & \multicolumn{1}{c|}{}                         & 92.86                                  & \multicolumn{1}{c|}{}                         & \cellcolor[HTML]{C0C0C0}90.64        \\
&                                                                             & C$\&$W                   & \multicolumn{1}{c|}{}                         & 96.18                                  & \multicolumn{1}{c|}{}                         & 98.26                                  & \multicolumn{1}{c|}{}                         & \cellcolor[HTML]{C0C0C0}83.68        &                                                                                   &                                                                             & C$\&$W                   & \multicolumn{1}{c|}{}                         & 45.55                                  & \multicolumn{1}{c|}{}                         & 99.72                                  & \multicolumn{1}{c|}{}                         & \cellcolor[HTML]{C0C0C0}100          \\
&                                                                             & IDSGAN                   & \multicolumn{1}{c|}{}                         & 71.75                                  & \multicolumn{1}{c|}{}                         & 93.05                                  & \multicolumn{1}{c|}{}                         & \cellcolor[HTML]{C0C0C0}69.44        &                                                                                   &                                                                             & IDSGAN                   & \multicolumn{1}{c|}{}                         & 49.83                                  & \multicolumn{1}{c|}{}                         & 80.04                                  & \multicolumn{1}{c|}{}                         & \cellcolor[HTML]{C0C0C0}82.85        \\
&                                                                             & J-Attack                 & \multicolumn{1}{c|}{}                         & 100                                    & \multicolumn{1}{c|}{}                         & 98.61                                  & \multicolumn{1}{c|}{}                         & \cellcolor[HTML]{C0C0C0}96.29        &                                                                                   &                                                                             & J-Attack                 & \multicolumn{1}{c|}{}                         & 43.1                                   & \multicolumn{1}{c|}{}                         & 32.91                                  & \multicolumn{1}{c|}{}                         & \cellcolor[HTML]{C0C0C0}35.62        \\
\multirow{-15}{*}{Web Attack}                                                    & \multirow{-5}{*}{GRU}                                                       & TEAM(ours)               & \multicolumn{1}{c|}{\multirow{-15}{*}{75.93}} & \textbf{100}                           & \multicolumn{1}{c|}{\multirow{-15}{*}{61.35}} & \textbf{100}                           & \multicolumn{1}{c|}{\multirow{-15}{*}{20}}    & \cellcolor[HTML]{C0C0C0}\textbf{100} & \multirow{-15}{*}{DDos}                                                           & \multirow{-5}{*}{GRU}                                                       & TEAM(ours)               & \multicolumn{1}{c|}{\multirow{-15}{*}{0.04}}  & \textbf{100}                           & \multicolumn{1}{c|}{\multirow{-15}{*}{0.58}}  & \textbf{100}                           & \multicolumn{1}{c|}{\multirow{-15}{*}{0.04}}  & \cellcolor[HTML]{C0C0C0}\textbf{100} \\ \hline
\end{tabular}}
\label{tab:CICAEattack}
\end{table*}

The adversarial attack effect of the J-Attack method on the NSL-KDD dataset is generally poor, and there is even a situation where the ASR is 0. This is because the adversarial attack of J-Attack is based on feature selection and is implemented by directly adding perturbations to the selected features. However, the NSL-KDD dataset itself has fewer statistical features, and statistical features are usually an important component of NIDS detection under deep learning. The lack of statistical features makes it difficult for J-Attack, a method that directly adds perturbations to selected features, to achieve accurate adversarial attacks. Furthermore, the selection of non-functional feature areas narrows the feature selection range of J-Attack and weakens the adversarial effect of J-Attack. Meanwhile, the information difference between PD-RNN and the target RNN in the past moments also increases the uncertainty of the adversarial effect of J-Attack, making it perform poorly on the NSL-KDD dataset.

In addition, in U2R$\&$R2L and Probe attacks, the PGD method uses multi-step iterations of small distance reverse gradients to implement AE attacks. Therefore, most AEs will not have a one-time iteration reverse gradient distance that is too large, causing AEs to approach the decision boundary on the other side of the attack target, making the AEs generated by the PGD method is relatively less affected by the difference weight distribution between PD-RNN and target RNN. Meanwhile, the U2R$\&$R2L attack itself basically does not have temporal, and the Probe attack has low temporal. Therefore, we believe that the AE generated by the PGD method on the U2R$\&$R2L and Probe attack categories is less affected by time interference. This is also the reason why the PGD method generates a part of AE white-box attack success rates and black-box transferability on the U2R$\&$R2L and Probe attack categories and is close to our TEAM method.

It is worth noting that in this paper, NIDS misjudges U2R$\&$R2L attacks as normal traffic with a high probability. This is because in network attacks, U2R$\&$R2L attacks are usually hidden in normal traffic, do not have high temporal, and have similar characteristics to normal traffic. Therefore, it is normal for NIDS to have a high probability of misjudging U2R$\&$R2L attacks as normal traffic.

\subsubsection{Adversarial Attacks on RNNs Using the CIC-IDS2017 Dataset}
In this section, we further verify the effectiveness of our proposed TEAM method on the CIC-IDS2017 dataset. The experimental results are shown in Table~\ref{tab:CICAEattack}.

\begin{table}[t]
\setlength{\arrayrulewidth}{0.3mm}
\caption{Time dilation ablation experiment in TEAM on NSL-KDD dataset (gray background represents white-box attacks, white background represents black-box transferability attacks). }
\renewcommand
\arraystretch{1.2}
\resizebox{\columnwidth}{!}{
\begin{tabular}{c|c|c|cc|cc|cc}
\Xhline{0.8pt}
 &                                                                          &                          & \multicolumn{2}{c|}{\begin{tabular}[c]{@{}c@{}}Original\\ RNN\end{tabular}}                         & \multicolumn{2}{c|}{LSTM}                                                             & \multicolumn{2}{c}{GRU}                                                               \\
\multirow{-2}{*}{\begin{tabular}[c]{@{}c@{}}Network\\ traffic type\end{tabular}} & \multirow{-2}{*}{\begin{tabular}[c]{@{}c@{}}IDS\\ Model\end{tabular}}    & \multirow{-2}{*}{Method} & MAR(\%)                                      & ASR(\%)                                              & MAR(\%)                                      & ASR(\%)                                & MAR(\%)                                      & ASR(\%)                                \\ \hline
&                                                                          & (W/O)TD                  & \multicolumn{1}{c|}{}                        & \cellcolor[HTML]{C0C0C0}{\color[HTML]{333333} 68.59} & \multicolumn{1}{c|}{}                        & 99.85                                  & \multicolumn{1}{c|}{}                        & 39.48                                  \\
 & \multirow{-2}{*}{\begin{tabular}[c]{@{}c@{}}Original\\ RNN\end{tabular}} & TEAM                     & \multicolumn{1}{c|}{}                        & \cellcolor[HTML]{C0C0C0}\textbf{100}                 & \multicolumn{1}{c|}{}                        & \textbf{100}                           & \multicolumn{1}{c|}{}                        & \textbf{100}                           \\ 
 \hhline{~|-|-|~|-|~|-|~|-}
&                                                                          & (W/O)TD                  & \multicolumn{1}{c|}{}                        & 100                                                  & \multicolumn{1}{c|}{}                        & \cellcolor[HTML]{C0C0C0}100            & \multicolumn{1}{c|}{}                        & 100                                    \\
    & \multirow{-2}{*}{LSTM}                                                   & TEAM                     & \multicolumn{1}{c|}{}                        & \textbf{100}                                         & \multicolumn{1}{c|}{}                        & \cellcolor[HTML]{C0C0C0}\textbf{100}   & \multicolumn{1}{c|}{}                        & \textbf{100}                           \\ 
     \hhline{~|-|-|~|-|~|-|~|-}
&                                                                          & (W/O)TD                  & \multicolumn{1}{c|}{}                        & 72.17                                                & \multicolumn{1}{c|}{}                        & 100                                    & \multicolumn{1}{c|}{}                        & \cellcolor[HTML]{C0C0C0}59             \\
\multirow{-6}{*}{Dos}                                                            & \multirow{-2}{*}{GRU}                                                    & TEAM                     & \multicolumn{1}{c|}{\multirow{-6}{*}{22.86}} & \textbf{100}                                         & \multicolumn{1}{c|}{\multirow{-6}{*}{23.87}} & \textbf{100}                           & \multicolumn{1}{c|}{\multirow{-6}{*}{21.7}}  & \cellcolor[HTML]{C0C0C0}\textbf{100}   \\ \hline
 &                                                                          & (W/O)TD                  & \multicolumn{1}{c|}{}                        & \cellcolor[HTML]{C0C0C0}99.32                        & \multicolumn{1}{c|}{}                        & 90.70                                  & \multicolumn{1}{c|}{}                        & 53.11                                  \\
 & \multirow{-2}{*}{\begin{tabular}[c]{@{}c@{}}Original\\ RNN\end{tabular}} & TEAM                     & \multicolumn{1}{c|}{}                        & \cellcolor[HTML]{C0C0C0}\textbf{99.77}               & \multicolumn{1}{c|}{}                        & \textbf{99.81}                         & \multicolumn{1}{c|}{}                        & \textbf{97.65}                         \\
  \hhline{~|-|-|~|-|~|-|~|-}
 &                                                                          & (W/O)TD                  & \multicolumn{1}{c|}{}                        & 99.59                                                & \multicolumn{1}{c|}{}                        & \cellcolor[HTML]{C0C0C0}99.77          & \multicolumn{1}{c|}{}                        & 99.54                                  \\
& \multirow{-2}{*}{LSTM}                                                   & TEAM                     & \multicolumn{1}{c|}{}                        & \textbf{99.81}                                       & \multicolumn{1}{c|}{}                        & \cellcolor[HTML]{C0C0C0}\textbf{99.86} & \multicolumn{1}{c|}{}                        & \textbf{99.81}                         \\ 
 \hhline{~|-|-|~|-|~|-|~|-}
&                                                                          & (W/O)TD                  & \multicolumn{1}{c|}{}                        & 99.72                                                & \multicolumn{1}{c|}{}                        & 99.63                                  & \multicolumn{1}{c|}{}                        & \cellcolor[HTML]{C0C0C0}97.78          \\
\multirow{-6}{*}{U2R$\&$R2L}                                                     & \multirow{-2}{*}{GRU}                                                    & TEAM                     & \multicolumn{1}{c|}{\multirow{-6}{*}{91.29}} & \textbf{99.77}                                       & \multicolumn{1}{c|}{\multirow{-6}{*}{91.97}} & \textbf{99.9}                          & \multicolumn{1}{c|}{\multirow{-6}{*}{91.09}} & \cellcolor[HTML]{C0C0C0}\textbf{99.86} \\ \hline
&                                                                          & (W/O)TD                  & \multicolumn{1}{c|}{}                        & \cellcolor[HTML]{C0C0C0}99.74                        & \multicolumn{1}{c|}{}                        & 100                                    & \multicolumn{1}{c|}{}                        & 100                                    \\
& \multirow{-2}{*}{\begin{tabular}[c]{@{}c@{}}Original\\ RNN\end{tabular}} & TEAM                     & \multicolumn{1}{c|}{}                        & \cellcolor[HTML]{C0C0C0}\textbf{100}                 & \multicolumn{1}{c|}{}                        & \textbf{100}                           & \multicolumn{1}{c|}{}                        & \textbf{100}                           \\ 
 \hhline{~|-|-|~|-|~|-|~|-}
 &                                                                          & (W/O)TD                  & \multicolumn{1}{c|}{}                        & 99.77                                                & \multicolumn{1}{c|}{}                        & \cellcolor[HTML]{C0C0C0}100            & \multicolumn{1}{c|}{}                        & 96.96                                  \\  & \multirow{-2}{*}{LSTM}                                                   & TEAM                     & \multicolumn{1}{c|}{}                        & \textbf{99.94}                                       & \multicolumn{1}{c|}{}                        & \cellcolor[HTML]{C0C0C0}\textbf{100}   & \multicolumn{1}{c|}{}                        & \textbf{100}                           \\ 
  \hhline{~|-|-|~|-|~|-|~|-}
&                                                                          & (W/O)TD                  & \multicolumn{1}{c|}{}                        & 87.3                                                 & \multicolumn{1}{c|}{}                        & 35.09                                  & \multicolumn{1}{c|}{}                        & \cellcolor[HTML]{C0C0C0}62.03          \\
\multirow{-6}{*}{Probe}                                                          & \multirow{-2}{*}{GRU}                                                    & TEAM                     & \multicolumn{1}{c|}{\multirow{-6}{*}{39.87}} & \textbf{100}                                         & \multicolumn{1}{c|}{\multirow{-6}{*}{27.27}} & \textbf{100}                           & \multicolumn{1}{c|}{\multirow{-6}{*}{36.65}} & \cellcolor[HTML]{C0C0C0}\textbf{100}   \\ \Xhline{0.8pt}
\end{tabular}}
\label{tab:wo}
\end{table}

From Table~\ref{tab:CICAEattack}, we can see that the proposed TEAM method shows excellent attack effects in all types of attacks compared with the existing four adversarial attack methods on the CIC-IDS2017 dataset, and is only slightly inferior to the IDSGAN method in the LSTM white-box attack of the Patator attack category. The reason is that the temporal of Patator, Web Attack and Infiltration$\&$Botnet is lower than that of Dos, Portscan and DDos attacks, and is less affected by temporal. Therefore, the C$\&$W, IDSGAN and J-Attack methods are less affected by the temporal difference between PD-RNN and the target RNN in these three types of attacks, and their adversarial attacks have good effects. In addition, we can see that the C$\&$W method has a good adversarial attack effect on LSTM and GRU of Dos, Portscan and DDos, but the effect of adversarial attacks on the original RNN is extremely poor. We believe that this is because the C$\&$W method directly optimizes the decision boundary of the classifier, making the adversarial sample as far away from the classification boundary as possible and close to the decision center of the target category. At the same time, LSTM and GRU can better control the flow of past information through the gating mechanism compared with the original RNN, thereby reducing the impact of past information on current information. Therefore, the C$\&$W method has achieved good results in Dos, Portscan and DDos LSTM and GRU adversarial attacks, but it is very susceptible to the impact of past information on original RNN, resulting in extremely poor results. The PGD method uses multi-step iterations of small distance reverse gradients to implement AE attacks. Therefore, the AE generated by it is usually close to the decision boundary and is easily disturbed by the difference in past information, resulting in slight  inferior adversarial attack effects on all attack types. The TEAM we proposed can use the time dilation method to make the weight distribution between PD-RNN and the target RNN closer, thereby achieving better adversarial attacks at any temporal strength. However, it should be admitted that this time dilation method is difficult to make the weight distribution between PD-RNN and the target RNN completely fit. In the Patator attack type that is inferior affected by temporal, since the time dilation method needs to take into account the effectiveness of various RNN attacks, this makes the goal of the time dilation weight distribution fitting more inclined to the universality of various RNN adversarial attacks, rather than specific RNN attacks. Therefore, compared with the white-box adversarial attack on LSTM in IDSGAN, the attack success rate is slightly lower.

\begin{table*}[]
\setlength{\arrayrulewidth}{0.3mm}
\caption{Time dilation ablation experiment in TEAM on CIC-IDS2017 dataset (gray background represents white-box attacks, white background represents black-box transferability attacks). }
\renewcommand
\arraystretch{1.2}  
\resizebox{\linewidth}{!}{
\begin{tabular}{c|c|c|cc|cc|cc|c|c|c|cc|cc|cc}
\hline
&                                                                          &                          & \multicolumn{2}{c|}{Original RNN}                                                     & \multicolumn{2}{c|}{LSTM}                                                             & \multicolumn{2}{c|}{GRU}                                                            &                                                                                     &                                                                          &                          & \multicolumn{2}{c|}{Original RNN}                                                     & \multicolumn{2}{c|}{LSTM}                                                             & \multicolumn{2}{c}{GRU}                                                             \\
\hhline{~|~|~|-|-|-|-|-|-|~|~|~|-|-|-|-|-|-}
\multirow{-2}{*}{\begin{tabular}[c]{@{}c@{}}Network\\ traffic type\end{tabular}} & \multirow{-2}{*}{\begin{tabular}[c]{@{}c@{}}IDS\\ Model\end{tabular}}    & \multirow{-2}{*}{Method} & \multicolumn{1}{c|}{MAR(\%)}                 & ASR(\%)                                & \multicolumn{1}{c|}{MAR(\%)}                 & ASR(\%)                                & \multicolumn{1}{c|}{MAR(\%)}                 & ASR(\%)                              & \multirow{-2}{*}{\begin{tabular}[c]{@{}c@{}}Network\\ traffic type\end{tabular}}    & \multirow{-2}{*}{\begin{tabular}[c]{@{}c@{}}IDS\\ Model\end{tabular}}    & \multirow{-2}{*}{Method} & \multicolumn{1}{c|}{MAR(\%)}                 & ASR(\%)                                & \multicolumn{1}{c|}{MAR(\%)}                 & ASR(\%)                                & \multicolumn{1}{c|}{ASR(\%)}                 & MAR(\%)                              \\ \hline
&                                                                          & (W/O)TD                  & \multicolumn{1}{c|}{}                        & \cellcolor[HTML]{C0C0C0}99.21          & \multicolumn{1}{c|}{}                        & 89.22                                  & \multicolumn{1}{c|}{}                        & 100                                  &                                                                                     &                                                                          & (W/O)TD                  & \multicolumn{1}{c|}{}                        & \cellcolor[HTML]{C0C0C0}67.82          & \multicolumn{1}{c|}{}                        & 99.22                                  & \multicolumn{1}{c|}{}                        & 100                                  \\
& \multirow{-2}{*}{\begin{tabular}[c]{@{}c@{}}Original\\ RNN\end{tabular}} & TEAM                     & \multicolumn{1}{c|}{}                        & \cellcolor[HTML]{C0C0C0}\textbf{99.97} & \multicolumn{1}{c|}{}                        & \textbf{99.8}                          & \multicolumn{1}{c|}{}                        & \textbf{100}                         &                                                                                     & \multirow{-2}{*}{\begin{tabular}[c]{@{}c@{}}Original\\ RNN\end{tabular}} & TEAM                     & \multicolumn{1}{c|}{}                        & \cellcolor[HTML]{C0C0C0}\textbf{97.67} & \multicolumn{1}{c|}{}                        & \textbf{100}                           & \multicolumn{1}{c|}{}                        & \textbf{100}                         \\
\hhline{~|-|-|~|-|~|-|~|-|~|-|-|~|-|~|-|~|-}
&                                                                          & (W/O)TD                  & \multicolumn{1}{c|}{}                        & 100                                    & \multicolumn{1}{c|}{}                        & \cellcolor[HTML]{C0C0C0}65.2           & \multicolumn{1}{c|}{}                        & 100                                  &                                                                                     &                                                                          & (W/O)TD                  & \multicolumn{1}{c|}{}                        & 62.79                                  & \multicolumn{1}{c|}{}                        & \cellcolor[HTML]{C0C0C0}94.57          & \multicolumn{1}{c|}{}                        & 100                                  \\
& \multirow{-2}{*}{LSTM}                                                   & TEAM                     & \multicolumn{1}{c|}{}                        & \textbf{100}                           & \multicolumn{1}{c|}{}                        & \cellcolor[HTML]{C0C0C0}\textbf{99.94} & \multicolumn{1}{c|}{}                        & \textbf{100}                         &                                                                                     & \multirow{-2}{*}{LSTM}                                                   & TEAM                     & \multicolumn{1}{c|}{}                        & \textbf{98.44}                         & \multicolumn{1}{c|}{}                        & \cellcolor[HTML]{C0C0C0}\textbf{100}   & \multicolumn{1}{c|}{}                        & \textbf{100}                         \\
\hhline{~|-|-|~|-|~|-|~|-|~|-|-|~|-|~|-|~|-}
&                                                                          & (W/O)TD                  & \multicolumn{1}{c|}{}                        & 100                                    & \multicolumn{1}{c|}{}                        & 97.42                                  & \multicolumn{1}{c|}{}                        & {\color[HTML]{000000} 100}           &                                                                                     &                                                                          & (W/O)TD                  & \multicolumn{1}{c|}{}                        & 58.91                                  & \multicolumn{1}{c|}{}                        & 98.83                                  & \multicolumn{1}{c|}{}                        & \cellcolor[HTML]{C0C0C0}100          \\
\multirow{-6}{*}{Dos}                                                            & \multirow{-2}{*}{GRU}                                                    & TEAM                     & \multicolumn{1}{c|}{\multirow{-6}{*}{24.09}} & \textbf{100}                           & \multicolumn{1}{c|}{\multirow{-6}{*}{9.41}}  & \textbf{99.97}                         & \multicolumn{1}{c|}{\multirow{-6}{*}{2.72}}  & {\color[HTML]{000000} \textbf{100}}  & \multirow{-6}{*}{\begin{tabular}[c]{@{}c@{}}Infiltration\\ $\&$Botnet\end{tabular}} & \multirow{-2}{*}{GRU}                                                    & TEAM                     & \multicolumn{1}{c|}{\multirow{-6}{*}{30.52}} & \textbf{100}                           & \multicolumn{1}{c|}{\multirow{-6}{*}{63.08}} & \textbf{100}                          & \multicolumn{1}{c|}{\multirow{-6}{*}{55.52}} & \cellcolor[HTML]{C0C0C0}\textbf{100} \\ \hline
&                                                                          & (W/O)TD                  & \multicolumn{1}{c|}{}                        & \cellcolor[HTML]{C0C0C0}100            & \multicolumn{1}{c|}{}                        & 99.3                                   & \multicolumn{1}{c|}{}                        & 100                                  &                                                                                     &                                                                          & (W/O)TD                  & \multicolumn{1}{c|}{}                        & \cellcolor[HTML]{C0C0C0}99.81          & \multicolumn{1}{c|}{}                        & 91.79                                  & \multicolumn{1}{c|}{}                        & 100                                  \\
& \multirow{-2}{*}{\begin{tabular}[c]{@{}c@{}}Original\\ RNN\end{tabular}} & TEAM                     & \multicolumn{1}{c|}{}                        & \cellcolor[HTML]{C0C0C0}\textbf{100}   & \multicolumn{1}{c|}{}                        & \textbf{100}                           & \multicolumn{1}{c|}{}                        & \textbf{100}                         &                                                                                     & \multirow{-2}{*}{\begin{tabular}[c]{@{}c@{}}Original\\ RNN\end{tabular}} & TEAM                     & \multicolumn{1}{c|}{}                        & \cellcolor[HTML]{C0C0C0}\textbf{100}   & \multicolumn{1}{c|}{}                        & \textbf{99.9}                          & \multicolumn{1}{c|}{}                        & \textbf{100}                         \\
\hhline{~|-|-|~|-|~|-|~|-|~|-|-|~|-|~|-|~|-}
&                                                                          & (W/O)TD                  & \multicolumn{1}{c|}{}                        & 100                                    & \multicolumn{1}{c|}{}                        & \cellcolor[HTML]{C0C0C0}97.74          & \multicolumn{1}{c|}{}                        & 100                                  &                                                                                     &                                                                          & (W/O)TD                  & \multicolumn{1}{c|}{}                        & 61                                     & \multicolumn{1}{c|}{}                        & \cellcolor[HTML]{C0C0C0}98.25          & \multicolumn{1}{c|}{}                        & 100                                  \\
& \multirow{-2}{*}{LSTM}                                                   & TEAM                     & \multicolumn{1}{c|}{}                        & \textbf{100}                           & \multicolumn{1}{c|}{}                        & \cellcolor[HTML]{C0C0C0}\textbf{99.82} & \multicolumn{1}{c|}{}                        & \textbf{100}                         &                                                                                     & \multirow{-2}{*}{LSTM}                                                   & TEAM                     & \multicolumn{1}{c|}{}                        & \textbf{96.68}                         & \multicolumn{1}{c|}{}                        & \cellcolor[HTML]{C0C0C0}\textbf{99.93} & \multicolumn{1}{c|}{}                        & \textbf{100}                         \\
\hhline{~|-|-|~|-|~|-|~|-|~|-|-|~|-|~|-|~|-}
&                                                                          & (W/O)TD                  & \multicolumn{1}{c|}{}                        & 100                                    & \multicolumn{1}{c|}{}                        & 100                                    & \multicolumn{1}{c|}{}                        & \cellcolor[HTML]{C0C0C0}100          &                                                                                     &                                                                          & (W/O)TD                  & \multicolumn{1}{c|}{}                        & 100                                    & \multicolumn{1}{c|}{}                        & 99.94                                  & \multicolumn{1}{c|}{}                        & \cellcolor[HTML]{C0C0C0}100          \\
\multirow{-6}{*}{Patator}                                                        & \multirow{-2}{*}{GRU}                                                    & TEAM                     & \multicolumn{1}{c|}{\multirow{-6}{*}{77.46}} & \textbf{100}                           & \multicolumn{1}{c|}{\multirow{-6}{*}{57.38}} & \textbf{100}                           & \multicolumn{1}{c|}{\multirow{-6}{*}{24.35}} & \cellcolor[HTML]{C0C0C0}\textbf{100} & \multirow{-6}{*}{Portscan}                                                          & \multirow{-2}{*}{GRU}                                                    & TEAM                     & \multicolumn{1}{c|}{\multirow{-6}{*}{0.5}}   & \textbf{100}                           & \multicolumn{1}{c|}{\multirow{-6}{*}{2.06}}  & \textbf{99.98}                         & \multicolumn{1}{c|}{\multirow{-6}{*}{9.4}}   & \cellcolor[HTML]{C0C0C0}\textbf{100} \\ \hline
&                                                                          & (W/O)TD                  & \multicolumn{1}{c|}{}                        & \cellcolor[HTML]{C0C0C0}100            & \multicolumn{1}{c|}{}                        & 96.29                                  & \multicolumn{1}{c|}{}                        & 87.96                                &                                                                                     &                                                                          & (W/O)TD                  & \multicolumn{1}{c|}{}                        & \cellcolor[HTML]{C0C0C0}63.69          & \multicolumn{1}{c|}{}                        & 99.98                                  & \multicolumn{1}{c|}{}                        & 100                                  \\
& \multirow{-2}{*}{\begin{tabular}[c]{@{}c@{}}Original\\ RNN\end{tabular}} & TEAM                     & \multicolumn{1}{c|}{}                        & \cellcolor[HTML]{C0C0C0}\textbf{100}   & \multicolumn{1}{c|}{}                        & \textbf{100}                           & \multicolumn{1}{c|}{}                        & \textbf{100}                         &                                                                                     & \multirow{-2}{*}{\begin{tabular}[c]{@{}c@{}}Original\\ RNN\end{tabular}} & TEAM                     & \multicolumn{1}{c|}{}                        & \cellcolor[HTML]{C0C0C0}\textbf{100}   & \multicolumn{1}{c|}{}                        & \textbf{100}                           & \multicolumn{1}{c|}{}                        & \textbf{100}                         \\
\hhline{~|-|-|~|-|~|-|~|-|~|-|-|~|-|~|-|~|-}
&                                                                          & (W/O)TD                  & \multicolumn{1}{c|}{}                        & 100                                    & \multicolumn{1}{c|}{}                        & \cellcolor[HTML]{C0C0C0}96.29          & \multicolumn{1}{c|}{}                        & 73.14                                &                                                                                     &                                                                          & (W/O)TD                  & \multicolumn{1}{c|}{}                        & 99.99                                  & \multicolumn{1}{c|}{}                        & \cellcolor[HTML]{C0C0C0}90.65          & \multicolumn{1}{c|}{}                        & 100                                  \\
& \multirow{-2}{*}{LSTM}                                                   & TEAM                     & \multicolumn{1}{c|}{}                        & \textbf{100}                           & \multicolumn{1}{c|}{}                        & \cellcolor[HTML]{C0C0C0}\textbf{100}   & \multicolumn{1}{c|}{}                        & \textbf{100}                         &                                                                                     & \multirow{-2}{*}{LSTM}                                                   & TEAM                     & \multicolumn{1}{c|}{}                        & \textbf{100}                           & \multicolumn{1}{c|}{}                        & \cellcolor[HTML]{C0C0C0}\textbf{100}   & \multicolumn{1}{c|}{}                        & \textbf{100}                         \\
\hhline{~|-|-|~|-|~|-|~|-|~|-|-|~|-|~|-|~|-}
&                                                                          & (W/O)TD                  & \multicolumn{1}{c|}{}                        & 100                                    & \multicolumn{1}{c|}{}                        & 100                                    & \multicolumn{1}{c|}{}                        & \cellcolor[HTML]{C0C0C0}79.16        &                                                                                     &                                                                          & (W/O)TD                  & \multicolumn{1}{c|}{}                        & 100                                    & \multicolumn{1}{c|}{}                        & 99.86                                  & \multicolumn{1}{c|}{}                        & \cellcolor[HTML]{C0C0C0}100          \\
\multirow{-6}{*}{Web Attack}                                                     & \multirow{-2}{*}{GRU}                                                    & TEAM                     & \multicolumn{1}{c|}{\multirow{-6}{*}{75.93}} & \textbf{100}                           & \multicolumn{1}{c|}{\multirow{-6}{*}{61.35}} & \textbf{100}                           & \multicolumn{1}{c|}{\multirow{-6}{*}{20}}    & \cellcolor[HTML]{C0C0C0}\textbf{100} & \multirow{-6}{*}{DDos}                                                              & \multirow{-2}{*}{GRU}                                                    & TEAM                     & \multicolumn{1}{c|}{\multirow{-6}{*}{0.04}}  & \textbf{100}                           & \multicolumn{1}{c|}{\multirow{-6}{*}{0.58}}  & \textbf{100}                           & \multicolumn{1}{c|}{\multirow{-6}{*}{0.04}}  & \cellcolor[HTML]{C0C0C0}\textbf{100} \\ \hline
\end{tabular}}
\label{tab:CICwo}
\end{table*}

In addition, we also observed that the success rate of adversarial attacks achieved by the four existing methods on the CIC-IDS2017 dataset is much higher than that on the NSL-KDD dataset. The reason is the CIC-IDS2017 dataset contains a large number of statistical features, which provide important judgment basis for NIDS detection methods based on deep learning. Meanwhile, statistical features usually have little impact on the functional implementation of network traffic. Even if the attack categories are different, in most cases, statistical features are usually retained as non-functional features. Adding adversarial perturbations to some statistical features reduces the difficulty of implementing adversarial attacks. Therefore, the ASR of the four existing methods on the CIC-IDS2017 dataset is higher than that on the NSL-KDD dataset.

\begin{table*}[]
\setlength{\arrayrulewidth}{0.3mm}
\caption{Next moment attack in RNN model. Among them, ASR1 represents the attack success rate of the OE on the RNN model at the first moment after the AEs, and ASR2 represents the attack success rate of the OE on the RNN model at the second moment after the AEs (gray background represents white-box attacks, white background represents black-box transferability attacks).}
\renewcommand
\arraystretch{1.2}  
\resizebox{\linewidth}{!}{
\begin{tabular}{c|c|c|c|cccc|cccc|cccc}
\hline
&                                                                                  &                                                                       &                          & \multicolumn{4}{c|}{Original RNN}       & \multicolumn{4}{c|}{LSTM}                                                                                                                                                        & \multicolumn{4}{c}{GRU}                                                                                                                                                         \\ \cline{5-16}
\multirow{-2}{*}{Dateset}                                                           & \multirow{-2}{*}{\begin{tabular}[c]{@{}c@{}}Network\\ traffic type\end{tabular}} & \multirow{-2}{*}{\begin{tabular}[c]{@{}c@{}}IDS\\ Model\end{tabular}} & \multirow{-2}{*}{Method} & \multicolumn{1}{c|}{MAR1(\%)}                & \multicolumn{1}{c|}{ASR1(\%)}                                             & \multicolumn{1}{c|}{MAR2(\%)}                & ASR2(\%)                      & \multicolumn{1}{c|}{MAR1(\%)}                & \multicolumn{1}{c|}{ASR1(\%)}                      & \multicolumn{1}{c|}{MAR2(\%)}                & ASR2(\%)                      & \multicolumn{1}{c|}{MAR1(\%)}                & \multicolumn{1}{c|}{ASR1(\%)}                      & \multicolumn{1}{c|}{MAR2(\%)}                & ASR2(\%)                      \\ \hline
&                                                                                  & Original RNN                                                          &                          & \multicolumn{1}{c|}{}                        & \multicolumn{1}{c|}{\cellcolor[HTML]{C0C0C0}{\color[HTML]{333333} 25.85}} & \multicolumn{1}{c|}{}                        & \cellcolor[HTML]{C0C0C0}21.78 & \multicolumn{1}{c|}{}                        & \multicolumn{1}{c|}{18.99}                         & \multicolumn{1}{c|}{}                        & 21.99                         & \multicolumn{1}{c|}{}                        & \multicolumn{1}{c|}{24.14}                         & \multicolumn{1}{c|}{}                        & 24.35                         \\ \hhline{~|~|-|~|~|-|~|-|~|-|~|-|~|-|~|-|}
&                                                                                  & LSTM                                                                  &                          & \multicolumn{1}{c|}{}                        & \multicolumn{1}{c|}{33.69}                                                & \multicolumn{1}{c|}{}                        & 21.45                         & \multicolumn{1}{c|}{}                        & \multicolumn{1}{c|}{\cellcolor[HTML]{C0C0C0}49.78} & \multicolumn{1}{c|}{}                        & \cellcolor[HTML]{C0C0C0}33.47 & \multicolumn{1}{c|}{}                        & \multicolumn{1}{c|}{49.78}                         & \multicolumn{1}{c|}{}                        & 33.47                         \\ \hhline{~|~|-|~|~|-|~|-|~|-|~|-|~|-|~|-|}
& \multirow{-3}{*}{Dos}                                                            & GRU                                                                   &                          & \multicolumn{1}{c|}{\multirow{-3}{*}{21.78}} & \multicolumn{1}{c|}{25.97}                                                & \multicolumn{1}{c|}{\multirow{-3}{*}{21.24}} & 23.71                         & \multicolumn{1}{c|}{\multirow{-3}{*}{21.78}} & \multicolumn{1}{c|}{30.25}                         & \multicolumn{1}{c|}{\multirow{-3}{*}{21.24}} & 28.96                         & \multicolumn{1}{c|}{\multirow{-3}{*}{23.28}} & \multicolumn{1}{c|}{\cellcolor[HTML]{C0C0C0}25.67} & \multicolumn{1}{c|}{\multirow{-3}{*}{22.85}} & \cellcolor[HTML]{C0C0C0}21.35 \\ \hhline{~|-|-|~|-|-|-|-|-|-|-|-|-|-|-|-|}
&                                                                                  & Original RNN                                                          &                          & \multicolumn{1}{c|}{}                        & \multicolumn{1}{c|}{\cellcolor[HTML]{C0C0C0}88.91}                        & \multicolumn{1}{c|}{}                        & \cellcolor[HTML]{C0C0C0}87.8  & \multicolumn{1}{c|}{}                        & \multicolumn{1}{c|}{94.3}                          & \multicolumn{1}{c|}{}                        & 89.15                         & \multicolumn{1}{c|}{}                        & \multicolumn{1}{c|}{91.59}                         & \multicolumn{1}{c|}{}                        & 89.7                          \\ \hhline{~|~|-|~|~|-|~|-|~|-|~|-|~|-|~|-|}
&                                                                                  & LSTM                                                                  &                          & \multicolumn{1}{c|}{}                        & \multicolumn{1}{c|}{94.57}                                                & \multicolumn{1}{c|}{}                        & 92.95                         & \multicolumn{1}{c|}{}                        & \multicolumn{1}{c|}{\cellcolor[HTML]{C0C0C0}94.03} & \multicolumn{1}{c|}{}                        & \cellcolor[HTML]{C0C0C0}92.68 & \multicolumn{1}{c|}{}                        & \multicolumn{1}{c|}{94.85}                         & \multicolumn{1}{c|}{}                        & 91.86                         \\ \hhline{~|~|-|~|~|-|~|-|~|-|~|-|~|-|~|-|}
& \multirow{-3}{*}{U2R$\&$R2L}                                                     & GRU                                                                   &                          & \multicolumn{1}{c|}{\multirow{-3}{*}{90.51}} & \multicolumn{1}{c|}{82.92}                                                & \multicolumn{1}{c|}{\multirow{-3}{*}{89.7}}  & 88.07                         & \multicolumn{1}{c|}{\multirow{-3}{*}{95.12}} & \multicolumn{1}{c|}{97.83}                         & \multicolumn{1}{c|}{\multirow{-3}{*}{94.85}} & 96.47                         & \multicolumn{1}{c|}{\multirow{-3}{*}{95.12}} & \multicolumn{1}{c|}{\cellcolor[HTML]{C0C0C0}93.49} & \multicolumn{1}{c|}{\multirow{-3}{*}{95.16}} & \cellcolor[HTML]{C0C0C0}90.78 \\ 
\hhline{~|-|-|~|-|-|-|-|-|-|-|-|-|-|-|-|}
&                                                                                  & Original RNN                                                          &                          & \multicolumn{1}{c|}{}                        & \multicolumn{1}{c|}{\cellcolor[HTML]{C0C0C0}38.74}                        & \multicolumn{1}{c|}{}                        & \cellcolor[HTML]{C0C0C0}41.72 & \multicolumn{1}{c|}{}                        & \multicolumn{1}{c|}{9.6}                           & \multicolumn{1}{c|}{}                        & 16.22                         & \multicolumn{1}{c|}{}                        & \multicolumn{1}{c|}{38.08}                         & \multicolumn{1}{c|}{}                        & 33.51                         \\ \hhline{~|~|-|~|~|-|~|-|~|-|~|-|~|-|~|-|}
&                                                                                  & LSTM                                                                  &                          & \multicolumn{1}{c|}{}                        & \multicolumn{1}{c|}{41.72}                                                & \multicolumn{1}{c|}{}                        & 39.4                          & \multicolumn{1}{c|}{}                        & \multicolumn{1}{c|}{\cellcolor[HTML]{C0C0C0}42.05} & \multicolumn{1}{c|}{}                        & \cellcolor[HTML]{C0C0C0}41.39 & \multicolumn{1}{c|}{}                        & \multicolumn{1}{c|}{42.05}                         & \multicolumn{1}{c|}{}                        & 39.73                         \\ \hhline{~|~|-|~|~|-|~|-|~|-|~|-|~|-|~|-|}
\multirow{-9}{*}{\begin{tabular}[c]{@{}c@{}}NSL-KDD\\ \\ Dataset\end{tabular}}      & \multirow{-3}{*}{Probe}                                                          & GRU                                                                   & \multirow{-9}{*}{TEAM}   & \multicolumn{1}{c|}{\multirow{-3}{*}{35.43}} & \multicolumn{1}{c|}{33.74}                                                & \multicolumn{1}{c|}{\multirow{-3}{*}{40.39}} & 40.72                         & \multicolumn{1}{c|}{\multirow{-3}{*}{25.82}} & \multicolumn{1}{c|}{4.63}                          & \multicolumn{1}{c|}{\multirow{-3}{*}{26.15}} & 20.86                         & \multicolumn{1}{c|}{\multirow{-3}{*}{33.11}} & \multicolumn{1}{c|}{\cellcolor[HTML]{C0C0C0}40.39} & \multicolumn{1}{c|}{\multirow{-3}{*}{33.44}} & \cellcolor[HTML]{C0C0C0}35.09 \\ \hline
&                                                                                  & Original RNN                                                          &                          & \multicolumn{1}{c|}{}                        & \multicolumn{1}{c|}{\cellcolor[HTML]{C0C0C0}85.66}                        & \multicolumn{1}{c|}{}                        & \cellcolor[HTML]{C0C0C0}87.45 & \multicolumn{1}{c|}{}                        & \multicolumn{1}{c|}{92.5}                          & \multicolumn{1}{c|}{}                        & 94.95                         & \multicolumn{1}{c|}{}                        & \multicolumn{1}{c|}{92.18}                         & \multicolumn{1}{c|}{}                        & 95.76                         \\ \hhline{~|~|-|~|~|-|~|-|~|-|~|-|~|-|~|-|}
&                                                                                  & LSTM                                                                  &                          & \multicolumn{1}{c|}{}                        & \multicolumn{1}{c|}{93.32}                                                & \multicolumn{1}{c|}{}                        & 95.6                          & \multicolumn{1}{c|}{}                        & \multicolumn{1}{c|}{\cellcolor[HTML]{C0C0C0}92.67} & \multicolumn{1}{c|}{}                        & \cellcolor[HTML]{C0C0C0}95.76 & \multicolumn{1}{c|}{}                        & \multicolumn{1}{c|}{93.64}                         & \multicolumn{1}{c|}{}                        & 96.09                         \\ \hhline{~|~|-|~|~|-|~|-|~|-|~|-|~|-|~|-|}
& \multirow{-3}{*}{Dos}                                                            & GRU                                                                   &                          & \multicolumn{1}{c|}{\multirow{-3}{*}{96.09}} & \multicolumn{1}{c|}{91.69}                                                & \multicolumn{1}{c|}{\multirow{-3}{*}{94.46}} & 93.64                         & \multicolumn{1}{c|}{\multirow{-3}{*}{94.62}} & \multicolumn{1}{c|}{96.9}                          & \multicolumn{1}{c|}{\multirow{-3}{*}{95.11}} & 96.57                         & \multicolumn{1}{c|}{\multirow{-3}{*}{92.5}}  & \multicolumn{1}{c|}{\cellcolor[HTML]{C0C0C0}97.71} & \multicolumn{1}{c|}{\multirow{-3}{*}{94.13}} & \cellcolor[HTML]{C0C0C0}98.04 \\ \hhline{~|-|-|~|-|-|-|-|-|-|-|-|-|-|-|-|}
&                                                                                  & Original RNN                                                          &                          & \multicolumn{1}{c|}{}                        & \multicolumn{1}{c|}{\cellcolor[HTML]{C0C0C0}{\color[HTML]{000000} 93.75}} & \multicolumn{1}{c|}{}                        & \cellcolor[HTML]{C0C0C0}100   & \multicolumn{1}{c|}{}                        & \multicolumn{1}{c|}{100}                           & \multicolumn{1}{c|}{}                        & 100                           & \multicolumn{1}{c|}{}                        & \multicolumn{1}{c|}{92.7}                          & \multicolumn{1}{c|}{}                        & 100                           \\ \hhline{~|~|-|~|~|-|~|-|~|-|~|-|~|-|~|-|}
&                                                                                  & LSTM                                                                  &                          & \multicolumn{1}{c|}{}                        & \multicolumn{1}{c|}{86.45}                                                & \multicolumn{1}{c|}{}                        & 95.83                         & \multicolumn{1}{c|}{}                        & \multicolumn{1}{c|}{\cellcolor[HTML]{C0C0C0}100}   & \multicolumn{1}{c|}{}                        & \cellcolor[HTML]{C0C0C0}100   & \multicolumn{1}{c|}{}                        & \multicolumn{1}{c|}{92.7}                          & \multicolumn{1}{c|}{}                        & 98.95                         \\ \hhline{~|~|-|~|~|-|~|-|~|-|~|-|~|-|~|-|}
& \multirow{-3}{*}{Patator}                                                        & GRU                                                                   &                          & \multicolumn{1}{c|}{\multirow{-3}{*}{94.79}} & \multicolumn{1}{c|}{95.83}                                                & \multicolumn{1}{c|}{\multirow{-3}{*}{98.95}} & 100                           & \multicolumn{1}{c|}{\multirow{-3}{*}{100}}   & \multicolumn{1}{c|}{95.83}                         & \multicolumn{1}{c|}{\multirow{-3}{*}{100}}   & 100                           & \multicolumn{1}{c|}{\multirow{-3}{*}{98.95}} & \multicolumn{1}{c|}{\cellcolor[HTML]{C0C0C0}92.7}  & \multicolumn{1}{c|}{\multirow{-3}{*}{98.95}} & \cellcolor[HTML]{C0C0C0}98.95 \\ \hhline{~|-|-|~|-|-|-|-|-|-|-|-|-|-|-|-|}
&                                                                                  & Original RNN                                                          &                          & \multicolumn{1}{c|}{}                        & \multicolumn{1}{c|}{\cellcolor[HTML]{C0C0C0}{\color[HTML]{000000} 100}}   & \multicolumn{1}{c|}{}                        & \cellcolor[HTML]{C0C0C0}94.44 & \multicolumn{1}{c|}{}                        & \multicolumn{1}{c|}{100}                           & \multicolumn{1}{c|}{}                        & 97.22                         & \multicolumn{1}{c|}{}                        & \multicolumn{1}{c|}{86.11}                         & \multicolumn{1}{c|}{}                        & 91.66                         \\ \hhline{~|~|-|~|~|-|~|-|~|-|~|-|~|-|~|-|}
&                                                                                  & LSTM                                                                  &                          & \multicolumn{1}{c|}{}                        & \multicolumn{1}{c|}{94.44}                                                & \multicolumn{1}{c|}{}                        & 100                           & \multicolumn{1}{c|}{}                        & \multicolumn{1}{c|}{\cellcolor[HTML]{C0C0C0}100}   & \multicolumn{1}{c|}{}                        & \cellcolor[HTML]{C0C0C0}100   & \multicolumn{1}{c|}{}                        & \multicolumn{1}{c|}{97.22}                         & \multicolumn{1}{c|}{}                        & 100                           \\ \hhline{~|~|-|~|~|-|~|-|~|-|~|-|~|-|~|-|}
& \multirow{-3}{*}{Web attack}                                                     & GRU                                                                   &                          & \multicolumn{1}{c|}{\multirow{-3}{*}{88.88}} & \multicolumn{1}{c|}{94.44}                                                & \multicolumn{1}{c|}{\multirow{-3}{*}{88.88}} & 97.22                         & \multicolumn{1}{c|}{\multirow{-3}{*}{94.44}} & \multicolumn{1}{c|}{100}                           & \multicolumn{1}{c|}{\multirow{-3}{*}{94.44}} & 100                           & \multicolumn{1}{c|}{\multirow{-3}{*}{94.44}} & \multicolumn{1}{c|}{\cellcolor[HTML]{C0C0C0}97.22} & \multicolumn{1}{c|}{\multirow{-3}{*}{86.11}} & \cellcolor[HTML]{C0C0C0}100   \\ \hhline{~|-|-|~|-|-|-|-|-|-|-|-|-|-|-|-|}
&                                                                                  & Original RNN                                                          &                          & \multicolumn{1}{c|}{}                        & \multicolumn{1}{c|}{\cellcolor[HTML]{C0C0C0}58.13}                        & \multicolumn{1}{c|}{}                        & \cellcolor[HTML]{C0C0C0}67.44 & \multicolumn{1}{c|}{}                        & \multicolumn{1}{c|}{100}                           & \multicolumn{1}{c|}{}                        & 100                           & \multicolumn{1}{c|}{}                        & \multicolumn{1}{c|}{86.04}                         & \multicolumn{1}{c|}{}                        & 100                           \\ \hhline{~|~|-|~|~|-|~|-|~|-|~|-|~|-|~|-|}
&                                                                                  & LSTM                                                                  &                          & \multicolumn{1}{c|}{}                        & \multicolumn{1}{c|}{55.81}                                                & \multicolumn{1}{c|}{}                        & 60.46                         & \multicolumn{1}{c|}{}                        & \multicolumn{1}{c|}{\cellcolor[HTML]{C0C0C0}100}   & \multicolumn{1}{c|}{}                        & \cellcolor[HTML]{C0C0C0}100   & \multicolumn{1}{c|}{}                        & \multicolumn{1}{c|}{86.04}                         & \multicolumn{1}{c|}{}                        & 100                           \\ \hhline{~|~|-|~|~|-|~|-|~|-|~|-|~|-|~|-|}
& \multirow{-3}{*}{Infiltration$\&$Botnet}                                            & GRU                                                                   &                          & \multicolumn{1}{c|}{\multirow{-3}{*}{74.41}} & \multicolumn{1}{c|}{62.79}                                                & \multicolumn{1}{c|}{\multirow{-3}{*}{79.06}} & 67.44                         & \multicolumn{1}{c|}{\multirow{-3}{*}{86.04}} & \multicolumn{1}{c|}{100}                           & \multicolumn{1}{c|}{\multirow{-3}{*}{93.02}} & 100                           & \multicolumn{1}{c|}{\multirow{-3}{*}{83.72}} & \multicolumn{1}{c|}{\cellcolor[HTML]{C0C0C0}95.34} & \multicolumn{1}{c|}{\multirow{-3}{*}{86.04}} & \cellcolor[HTML]{C0C0C0}95.34 \\ \hhline{~|-|-|~|-|-|-|-|-|-|-|-|-|-|-|-|}
&                                                                                  & Original RNN                                                          &                          & \multicolumn{1}{c|}{}                        & \multicolumn{1}{c|}{\cellcolor[HTML]{C0C0C0}2.7}                          & \multicolumn{1}{c|}{}                        & \cellcolor[HTML]{C0C0C0}2.5   & \multicolumn{1}{c|}{}                        & \multicolumn{1}{c|}{99.29}                         & \multicolumn{1}{c|}{}                        & 99.92                         & \multicolumn{1}{c|}{}                        & \multicolumn{1}{c|}{99.86}                         & \multicolumn{1}{c|}{}                        & 95.3                          \\ \hhline{~|~|-|~|~|-|~|-|~|-|~|-|~|-|~|-|}
&                                                                                  & LSTM                                                                  &                          & \multicolumn{1}{c|}{}                        & \multicolumn{1}{c|}{2.04}                                                 & \multicolumn{1}{c|}{}                        & 1.62                          & \multicolumn{1}{c|}{}                        & \multicolumn{1}{c|}{\cellcolor[HTML]{C0C0C0}88.99} & \multicolumn{1}{c|}{}                        & \cellcolor[HTML]{C0C0C0}97.03 & \multicolumn{1}{c|}{}                        & \multicolumn{1}{c|}{99.23}                         & \multicolumn{1}{c|}{}                        & 78.94                         \\ \hhline{~|~|-|~|~|-|~|-|~|-|~|-|~|-|~|-|}
& \multirow{-3}{*}{Portscan}                                                       & GRU                                                                   &                          & \multicolumn{1}{c|}{\multirow{-3}{*}{2.28}}  & \multicolumn{1}{c|}{2.15}                                                 & \multicolumn{1}{c|}{\multirow{-3}{*}{1.81}}  & 2.04                          & \multicolumn{1}{c|}{\multirow{-3}{*}{4.35}}  & \multicolumn{1}{c|}{98.89}                         & \multicolumn{1}{c|}{\multirow{-3}{*}{4.17}}  & 99.89                         & \multicolumn{1}{c|}{\multirow{-3}{*}{4.72}}  & \multicolumn{1}{c|}{\cellcolor[HTML]{C0C0C0}99.73} & \multicolumn{1}{c|}{\multirow{-3}{*}{4.48}}  & \cellcolor[HTML]{C0C0C0}92.72 \\ \hhline{~|-|-|~|-|-|-|-|-|-|-|-|-|-|-|-|}
&                                                                                  & Original RNN                                                          &                          & \multicolumn{1}{c|}{}                        & \multicolumn{1}{c|}{\cellcolor[HTML]{C0C0C0}35.25}                        & \multicolumn{1}{c|}{}                        & \cellcolor[HTML]{C0C0C0}7.07  & \multicolumn{1}{c|}{}                        & \multicolumn{1}{c|}{86.25}                         & \multicolumn{1}{c|}{}                        & 64.27                         & \multicolumn{1}{c|}{}                        & \multicolumn{1}{c|}{94.96}                         & \multicolumn{1}{c|}{}                        & 79.27                         \\ \hhline{~|~|-|~|~|-|~|-|~|-|~|-|~|-|~|-|}
&                                                                                  & LSTM                                                                  &                          & \multicolumn{1}{c|}{}                        & \multicolumn{1}{c|}{42.01}                                                & \multicolumn{1}{c|}{}                        & 18.63                         & \multicolumn{1}{c|}{}                        & \multicolumn{1}{c|}{\cellcolor[HTML]{C0C0C0}83.97} & \multicolumn{1}{c|}{}                        & \cellcolor[HTML]{C0C0C0}61.99 & \multicolumn{1}{c|}{}                        & \multicolumn{1}{c|}{95.24}                         & \multicolumn{1}{c|}{}                        & 78.43                         \\ \hhline{~|~|-|~|~|-|~|-|~|-|~|-|~|-|~|-|}
\multirow{-18}{*}{\begin{tabular}[c]{@{}c@{}}CIC-IDS2017\\ \\ Dataset\end{tabular}} & \multirow{-3}{*}{DDos}                                                           & GRU                                                                   & \multirow{-18}{*}{TEAM}  & \multicolumn{1}{c|}{\multirow{-3}{*}{0.04}}  & \multicolumn{1}{c|}{32.55}                                                & \multicolumn{1}{c|}{\multirow{-3}{*}{0.04}}  & 8.61                          & \multicolumn{1}{c|}{\multirow{-3}{*}{0.09}}  & \multicolumn{1}{c|}{74.28}                         & \multicolumn{1}{c|}{\multirow{-3}{*}{0.09}}  & 68.09                         & \multicolumn{1}{c|}{\multirow{-3}{*}{0}}     & \multicolumn{1}{c|}{\cellcolor[HTML]{C0C0C0}95.57} & \multicolumn{1}{c|}{\multirow{-3}{*}{0}}     & \cellcolor[HTML]{C0C0C0}80.2  \\ \hline
\end{tabular}}
\label{tab:nextforatt}
\end{table*}

\subsubsection{Validity Verification of Time Dilation Method}

In this section, we conduct ablation experiments on the time dilation method in our proposed TEAM method to verify the effectiveness of our proposed time dilation method.

As can be seen from Table~\ref{tab:wo} and Table~\ref{tab:CICwo}, the TEAM model that does not use the time dilation method performs poorly in white-box adversarial attacks and black-box transferability of the RNN model. The reason is there is a difference in retention of past moments between PD-RNN and target RNN during the training of the TEAM model. Therefore, the AE generated by TEAM will be affected by this difference, which in turn affects the corresponding adversarial attacks and transferability. Due to TEAM with time dilation method overlaps the weight distribution between PD-RNN and target RNN as much as possible, TEAM with time dilation method has better effect than TEAM without time dilation method. In addition, the attack on the next moment is also considered in the TEAM method, which intensifies AE's focus on temporal characteristics and amplifies the difference between PD-RNN and target RNN in data retention of past moments. Therefore, when the time dilation method is not used, the AE generated by our proposed TEAM is less effective in white-box adversarial attacks and black-box transferable attacks.

\subsubsection{Next Moment Attack}

In this section, we use experiments to verify the impact of AEs on subsequent OEs. That is, the attack at the next moment. Specifically, In the RNN model, the recombined data will change the structure of the data and affect the time distribution between the generated AEs, which will in turn affect the adversarial attacks of the AEs on the RNN model and the subsequent attacks. Since the experiment in this section aims to verify the concept of attack at the next moment, we did not conduct other comparative experiments. This concept is experimentally verified using only TEAM with time dilation method.

It can be seen from Table~\ref{tab:nextforatt} that AEs can effectively affect the detection of subsequent OEs by the RNN model.

In NSL-KDD dataset, our proposed method shows good attack and transferability performance on Dos and Probe. However, the performance on U2R$\&$R2L is poor, which is because Dos is a type of high-intensity continuous attack. This attack has strong temporal and is sensitive to the content of the past moment. Therefore, in the RNN model, a large amount of AEs data in the past moment can push the OEs at the next moment to cross the decision boundary of the model, thereby exacerbating the error of NIDS in misjudging the OEs of the Dos category as normal traffic. TEAM works better overall in next moment attacks because it retains a lot of past moment content.

In the Probe attack, although the temporal of the Probe attack is not as good as the temporal of the Dos attack type. However, the temporal  of this type of attack also makes Probe more sensitive to the content of the past moment. Therefore, the attack at the next moment will have similar analysis and effects on the Probe attack type as on the Dos attack type. We also observed that in the Probe attack type, some next moment attack have extremely low ASR. We observed the experimental results and found that the reason is that the past moment AEs generated by the Probe attack type have a greater impact on the temporal of current moment OEs, causing the past moment content to have an excessive impact on the current moment OEs, making the current moment OEs excessively cross the decision boundary, and it is judged as a Dos attack.

However, U2R$\&$R2L pays more attention to the content of the current moment, exhibit less temporal, and is not sensitive to the impact of time. Therefore, it is difficult for us to use the content of AEs in the past moment to push the OEs of U2R$\&$R2L at the next moment to cross the decision boundary of the model. In addition, the content of AEs in the past time may make the characteristic difference between U2R$\&$R2L and normal traffic more obvious, which is even beneficial to the detection of NIDS. This is why the next moment attack has poor effect on U2R$\&$R2L.

In CIC-IDS2017 dataset, we can see that before the next attack, continuous Dos and Patator attacks can already affect the RNN classifier, causing the classifier to greatly misjudge the Dos and Patator attack type traffic of the OE (\textit{i.e.}, with higher MSR1 and MSR2). This is because Dos data has a strong temporal nature, making it extremely susceptible to the influence of past information. Even if the previous moment is OE traffic, it can easily cause the subsequent Dos attack to cross the model decision boundary and be misjudged as normal traffic. Patator is because its statistical characteristics are usually very similar to those of normal traffic. Even if the previous moment is OE traffic, the influence of its information on the statistical characteristics of subsequent moments can easily cause the Patator attack type to cross the model's decision boundary and form a misjudgment.

When we perform the next moment attack, we can find that the next moment attack effect of Dos and Patator attack types is poor. This is because the traffic that was AE at the previous moment increases the influence of the past moment information on the subsequent moment, which in turn causes the next moment attack to cross the boundary (for example, Patator should have been identified as normal, but some subsequent moments OE crossed the decision boundary of normal and were identified as Dos). In addition, in other attack categories, we can clearly see that the AE in the past moment effectively affects the detection of OE attack types in the next moment and the next next moment by the NIDS based on RNN. It makes the samples that should have been detected as attacks by NIDS have a very high probability of being misjudged as normal examples, with the maximum probability even being 100\%. This greatly affects the security of NIDS based on RNN.

In addition, we can see that the next moment attack misjudgment rate on the CIC-IDS2017 dataset is generally higher than that on the NSL-KDD dataset. The reason is the CIC-IDS2017 dataset has more traffic features that are consistent with modern machine learning discrimination than NSL-KDD, such as more statistical features. This also indirectly improves the misjudgment rate of the NIDS model based on RNN.

Overall, the attack at the next moment increases the misjudgment rate of OE by NIDS. In Dos and Probe of NSL-KDD dataset, the increase of NIDS misjudgment rate even exceeds 2 times, and in the DDos of CIC-IDS2017, the misjudgment rate of NIDS increased from the original 0\% to an astonishing 95.57\%. The attacker only needs to create a small number of AEs to greatly affect subsequent OEs in the same time step and carry out further attacks. This is a very dangerous new problem for NIDS.

\section{Conclusion}

In this paper, we conducted the study of adversarial attacks on RNN models with time steps in NIDS. We first designed a new AEs generation method for RNN models called TEAM. Then, we used TEAM to generate AEs for adversarial attacks and next moment attacks on RNN model, revealing the potential connection between adversarial and time steps in RNN model. Finally, through a large number of experiments, we verified that the AEs generated by TEAM can be effectively used for adversarial attacks and next moment attacks on the RNN model in NIDS.

In the future, we will devote ourself to defense adversarial attacks and defense adversarial attacks at the next moment. Meanwhile, we will investigate how to defend against temporal adversarial attacks and next moment attacks caused by temporal in RNN-based NIDS, thereby enhancing the security of NIDS.

\newpage






\vfill

\end{document}